\documentclass[aps,prd,floats,epsf,superscriptaddress,nofootinbib,amsmath]{revtex4}
\usepackage{amsmath,amssymb}
\usepackage{graphicx}
\usepackage{dsfont}
\usepackage{slashed}
\usepackage{multirow}
\usepackage{color}
\usepackage{longtable}
\usepackage[colorlinks=true,citecolor=blue,linkcolor=blue,urlcolor=blue]{hyperref}
\usepackage{ulem}
\allowdisplaybreaks

\begin{document}

\title{Standard Model Effective Field Theory:\\
Integrating out Neutralinos and Charginos in the MSSM}

\author{Huayong Han}
\email[e-mail: ]{han@itp.ac.cn}
\affiliation{CAS Key Laboratory of Theoretical Physics, Institute of Theoretical Physics, Chinese Academy of Sciences, Beijing 100190, P. R. China}

\author{Ran Huo}
\email[e-mail: ]{huora@ucr.edu}
\affiliation{Department of Physics and Astronomy, University of California, Riverside, California 92521, USA}

\author{Minyuan Jiang}
\email[e-mail: ]{dg1522023@smail.nju.edu.cn}
\affiliation{Department of Physics, Nanjing University, 22 Hankou Road, Nanjing 210098, P. R. China}
\affiliation{CAS Key Laboratory of Theoretical Physics, Institute of Theoretical Physics, Chinese Academy of Sciences, Beijing 100190, P. R. China}

\author{Jing Shu}
\email[e-mail: ]{jshu@itp.ac.cn}
\affiliation{CAS Key Laboratory of Theoretical Physics, Institute of Theoretical Physics, Chinese Academy of Sciences, Beijing 100190, P. R. China}
\affiliation{School of Physical Sciences, University of Chinese Academy of Sciences, Beijing 100049, P. R. China}
\affiliation{CAS Center for Excellence in Particle Physics, Beijing 100049, P. R. China}
\affiliation{Center for High Energy Physics, Peking University, Beijing 100871, China}

\date{\today}

\begin{abstract}
We apply the covariant derivative expansion method to integrate out the neutralinos and charginos in the minimal supersymmetric Standard Model. The results are presented as  set of pure bosonic dimension-6 operators in the Standard Model effective field theory. Nontrivial chirality dependence in fermionic covariant derivative expansion are discussed carefully. The results are checked by computing the $h\gamma\gamma$ effective coupling and the electroweak oblique parameters using the Standard Model effective field theory with our effective operators and  direct loop calculation. In global fitting the proposed lepton collider constraint projections, special phenomenological emphasis is paid to the gaugino mass unification scenario ($M_2\simeq2M_1$) and anomaly mediation scenario ($M_1\simeq3.3M_2$). These results show that the precision measurement experiments in future lepton colliders will provide a very useful complementary job in probing the electroweakino sector, in particular, filling the gap of the soft lepton plus missing $E_T$ channel search left by the traditional collider, where the neutralino as the lightest supersymmetric particle is very degenerated with the next-to-lightest  chargino/neutralino.
\end{abstract}

\maketitle

\flushbottom

\section{Introduction}
\label{sec:introduction}

As an effective tool of connecting the new physics to the experimental observations, the Standard Model effective field theory (SMEFT) is recently being intensively developed and applied (refer to ~\cite{Henning:2014wua,Brivio:2017vri} for a review).
One advantage of the SMEFT is that it provides a model independent way when comparing with experimental results.
In the SMEFT, the beyond Standard Model (SM) effects are characterized by a series of high dimensional operators, which are constructed with the SM fields only and satisfy the SM $SU(3)_C\times SU(2)_L\times U(1)_Y$ symmetry.
Regarding the precision of the current and near future experiments, it is usually sufficient to consider only dimension-6 operators, the number of which is finite (59 when demanding baryon number conservation)~\cite{Buchmuller:1985jz,Grzadkowski:2010es}.
The experimental observables can be expressed by SM parameters and Wilson coefficients of high dimensional operators.
Then the results of experiments can be transformed into the constrains on these Wilson coefficients.
For theorists who have their favorite models and want to test them through experiments, the main work is to calculate those coefficients of the effective operators from the model.

To effectively match a new physics model to the SMEFT, the method of covariant derivative expansion (CDE), initially introduced in~\cite{Gaillard:1985uh,Chan:1986jq,Cheyette:1987qz}, is being well developed recently~\cite{Henning:2014wua,Drozd:2015rsp,Ellis:2016enq,Henning:2016lyp,Zhang:2016pja,delAguila:2016zcb,Fuentes-Martin:2016uol,Ellis:2017jns}. Comparing to the traditional way of Feynman diagram calculations, which is relatively tedious and not gauge invariant in the process, the CDE method can directly obtain the gauge-invariant operators in a much simpler way. Using this method, the effective operators have been calculated for various new physics models and the corresponding phenomenological implications were discussed~\cite{Huo:2015nka,Huo:2015exa,Chiang:2015ura,Henning:2014gca, Wells:2017vla,Cao:2017oez}.

A particularly important kind of the beyond SM models is the supersymmetric one, which provides elegant solutions to the naturalness problem, the dark matter candidates and the path to the grand unification. Applying the CDE method, people have already worked out the dimension-6 effective operators by integrating out the sfermions in the minimal supersymmetric standard model (MSSM) to one-loop order. In this work we consider another important sector of the MSSM, the neutralinos and charginos (or called the electroweakino sector). The electroweakino sector of the MSSM is of great significance in phenomenology, because the lightest neutralino is often considered as a very competitive dark matter candidate.
We calculate dimension-6 operators by integrating out the neutralinos and charginos with the CDE method. Since the original form of the electroweakino sector is expressed with two-component Weyl fermions, the procedure is not as straightforward as those in literature. We show in detail how to transform the Lagrangian to a desired form to apply the CDE and then obtain the effective operators with the universal one-loop effective action~\cite{Drozd:2015rsp}.
Our results can be used in a lot of phenomenological studies of the effects of the neutralinos and charginos, e.g.,  cosmology~\cite{Cao:2017oez,Huang:2016odd,Gan:2017mcv}  and collider indirect searches~\cite{Ellis:2017kfi,Ferreira:2016jea,Liu:2016gzs,Craig:2015wwr,Liu:2016idz}.
As an illustration, we apply our results in future lepton colliders, to see how the precise electroweak and Higgs observables  constrain the MSSM electroweakino sector.


The paper is organized as follows. Section \ref{sec:formalism} presents the key steps to apply the CDE formalism to the electroweakino secctor. The analytical cross-check against other calculation methods are displayed in section \ref{sec:results}, and the constraints on parameter space of MSSM are discussed in section \ref{sec:constraint}. We summarize and conclude in section \ref{sec:conclude}. Some discussions about the regularization options and the Wilson coefficients of the effective operators are presented in appendices.

\section{The Formulism and Input}
\label{sec:formalism}

Each of the two Higgs superfield has the same degree of freedom with the supersymmetrization of the left chiral SM lepton, and the gaugino is also of Majorana type and has the same degree of freedom with a Weyl fermion, so the natural way to write down the Lagrangian is in the two-component Weyl spinor form. Taking the left chiral form for example,~\cite{Martin:1997ns}
\begin{eqnarray}
\mathcal{L}_\text{MSSM} \supset &&i\widetilde{B}^\dagger\bar{\sigma}^{\mu}D_{\mu}\widetilde{B}
+i\widetilde{W}^\dagger\bar{\sigma}^{\mu}D_{\mu}\widetilde{W}
+i\widetilde{H}_u^\dagger\bar{\sigma}^{\mu}D_{\mu}\widetilde{H}_u
+i\widetilde{H}_d^\dagger\bar{\sigma}^{\mu}D_{\mu}\widetilde{H}_d \nonumber \\
&&-\frac{1}{2}(M_1\widetilde{B}\widetilde{B}+M_2\widetilde{W}\widetilde{W}+c.c)
-\mu(\widetilde{H}_u^T\epsilon\widetilde{H}_d+c.c) \nonumber \\
&&+[\frac{\sqrt{2}}{2}g(H_u^\dagger\sigma^a\widetilde{H}_u+H_d^\dagger\sigma^a\widetilde{H}_d)\widetilde{W}^a+ \frac{\sqrt{2}}{2}g'(H_u^\dagger\widetilde{H}_u-H_d^\dagger\widetilde{H}_d)\widetilde{B}+c.c],
\label{eq:lagrangian}
\end{eqnarray}
where $\tilde{H}_u = (\widetilde{H}_u^+,\, \widetilde{H}_u^0)^T$, $ \widetilde{H}_d = (\widetilde{H}_d^{0\ast},\, -\widetilde{H}_d^-)^T$ (so that the charge conjugation $\widetilde{H}_d^c= \epsilon \widetilde{H}_d^\ast$ has the same phase with $\widetilde{H}_u$), $ \widetilde{W} = (\widetilde{W}^1,\, \widetilde{W}^2,\,\widetilde{W}^3)^T$ and $\widetilde{B}$ are the higgsinos and electroweak gauginos, and $(\enspace)^T$ indicates transpose. Here the $\epsilon=-i\sigma^2=(^0_1~^{-1}_{~0})$ is the two dimensional antisymmetric matrix used to contract the two $SU(2)$ doublets $\widetilde{H}_u$ and $\widetilde{H}_d$, while other $\epsilon$ matrixes used to contract two Weyl spinors are not shown explicitly. And $\sigma^a$ are the Pauli matrices.

We follow the procedure of CDE method~\cite{Henning:2014wua} to integrate out the neutralinos and charginos in the MSSM at one loop level. Since the CDE directly works in the four-component Dirac spinors, in the following we will show the procedures of organizing Eq.~(\ref{eq:lagrangian}) into that form.
Because $\xi^\dag i\bar{\sigma}^\mu\partial_\mu\chi=\chi i\sigma^\mu\partial_\mu\xi^\dag$ (which follows  the identity $\xi^\dag\bar{\sigma}^\mu\chi=-\chi\sigma^\mu\xi^\dag$ and integration by parts), both the bino (wino) and the higgsino kinetic terms can be arranged in the Dirac form with a overall factor of $\frac{1}{2}$.
However, it is noted that the CDE formalism requires a unique covariant derivative in the basis where the BSM fields are organized with their SM representation components expanded, but this is not simply true for chiral fermions between different chiralities, if the fermions are in the fundamental representation of some non-Abelian gauge group. The most straightforward way to see it is through the identity $g\psi^\dag_i\bar{\sigma}^\mu\psi_jA_\mu^at^a_{ij}=-g\psi_j\sigma^\mu\psi^\dag_iA_\mu^at^a_{ij}$ in the two-component spinor basis, which indicates that an extra matrix transpose for the gauge group generator of the fundamental representation is needed\footnote{There is no such problem in the gaugino sector, due to the identity $f^{abc}=-f^{cba}$ for adjoint representation.}.

Another way to clarify such issue with the non-Abelian gauge field is in terms of the raising and lowering operators. For example, such $SU(2)_L$ gauge field on the doublet fundamental representation is usually chosen as $\frac{g}{\sqrt{2}}(^{~0}_{W^-}~^{W^+}_{~0})$, which is consistent with left chiral fermion ordering $(\widetilde{H}_u^+,\, \widetilde{H}_u^0,~\widetilde{H}_d^{0\ast},\, -\widetilde{H}_d^-)^T$ if it acts on the first two and last two components in a block diagonal way. For the right chiral fermion ordering to be consistently acted by the same gauge field, apparently the choice is $(\widetilde{H}_d^+,\, \widetilde{H}_d^0,~\widetilde{H}_u^{0\ast},\, -\widetilde{H}_u^-)^T$, or the overall charge conjugation of the left chiral one. Combining with the gaugino, the complete Dirac spinor form of the electroweakino is chosen as $\chi=(\chi^B,~\chi^1,~\chi^2,~\chi^3,~\chi^+,\chi^0,~\chi^{0\ast},-\chi^-)^T$ in the SM $SU(2)_L\times U(1)_Y$ component field basis, with each component of Dirac field written in the Van der Waerden form
\begin{align}
\chi^B=\left(\begin{array}{c}
\widetilde{B} \\
\widetilde{B}^\dag
\end{array}\right),~~\chi^1=\left(\begin{array}{c}
\widetilde{W}^1 \\
\widetilde{W}^{1\dag}
\end{array}\right),~~\chi^2=\left(\begin{array}{c}
\widetilde{W}^2 \\
\widetilde{W}^{2\dag}
\end{array}\right),~~\chi^3=\left(\begin{array}{c}
\widetilde{W}^3 \\
\widetilde{W}^{3\dag}
\end{array}\right),\\
\chi^+=\left(\begin{array}{c}
\widetilde{H}^+_u \\
\widetilde{H}^+_d
\end{array}\right),~~\chi^0=\left(\begin{array}{c}
\widetilde{H}^0_u \\
\widetilde{H}^0_d
\end{array}\right),~~\chi^{0\ast}=\left(\begin{array}{c}
\widetilde{H}^{0\ast}_d \\
\widetilde{H}^{0\ast}_u
\end{array}\right),~~\chi^-=\left(\begin{array}{c}
\widetilde{H}^-_d \\
\widetilde{H}^-_u
\end{array}\right).
\end{align}
Here we have suppressed the undotted and dotted spinor index, and for higgsino the physical electric charge is used instead of the (in)existence of a $\dag$.

Then we can apply the CDE. The fermionic new physics model path integral is formally written as
\begin{eqnarray}
e^{iS_\text{eff}} &=& \int D\bar{\chi} D\chi\exp\Bigg\{\frac{1}{2}\bar{\chi}(i\gamma^\mu D_\mu-M-U_LP_L-U_RP_R)\chi\Bigg\},
\label{eq:S_eff0}
\end{eqnarray}
where the $P_{L,R} = (1\mp \gamma_5)/2$ are left and right projection operators. We can get the SM field independent $M$ matrix and the dependent $U_L,U_R$ matrix
\begin{eqnarray}
M&=&\left(\begin{array}{cccccccc}
M_1 & & & & & & &  \\
 & M_2 & & & & & & \\
 & & M_2 & & & & & \\
 & & & M_2 & & & & \\
 & & & & \mu & & & \\
 & & & & & \mu & & \\
 & & & & & & \mu & \\
 & & & & & & & \mu
\end{array}\right),\\
U_L&=&\left(\begin{array}{cccccccc}
 & & & & -\frac{g'}{\sqrt{2}}H_u^- & -\frac{g'}{\sqrt{2}}H_u^{0\ast} & \frac{g'}{\sqrt{2}}H_d^0 & -\frac{g'}{\sqrt{2}}H_d^+ \\
 & & & & -\frac{g}{\sqrt{2}}H_u^{0\ast} & -\frac{g}{\sqrt{2}}H_u^- & \frac{g}{\sqrt{2}}H_d^+ & -\frac{g}{\sqrt{2}}H_d^0 \\
 & & & & -\frac{ig}{\sqrt{2}}H_u^{0\ast} & \frac{ig}{\sqrt{2}}H_u^- & \frac{ig}{\sqrt{2}}H_d^+ & \frac{ig}{\sqrt{2}}H_d^0 \\
 & & & & -\frac{g}{\sqrt{2}}H_u^- & \frac{g}{\sqrt{2}}H_u^{0\ast} & -\frac{g}{\sqrt{2}}H_d^0 & -\frac{g}{\sqrt{2}}H_d^+ \\
-\frac{g'}{\sqrt{2}}H_d^+ & -\frac{g}{\sqrt{2}}H_d^0 & \frac{ig}{\sqrt{2}}H_d^0 & -\frac{g}{\sqrt{2}}H_d^+ & & & & \\
-\frac{g'}{\sqrt{2}}H_d^0 & -\frac{g}{\sqrt{2}}H_d^+ & -\frac{ig}{\sqrt{2}}H_d^+ & \frac{g}{\sqrt{2}}H_d^0 & & & & \\
\frac{g'}{\sqrt{2}}H_u^{0\ast} & \frac{g}{\sqrt{2}}H_u^- & -\frac{ig}{\sqrt{2}}H_u^- & -\frac{g}{\sqrt{2}}H_u^{0\ast} & & & & \\
-\frac{g'}{\sqrt{2}}H_u^- & -\frac{g}{\sqrt{2}}H_u^{0\ast} & -\frac{ig}{\sqrt{2}}H_u^{0\ast} & -\frac{g}{\sqrt{2}}H_u^- & & & &
\end{array}\right),\\
U_R&=&U_L^\dag.
\end{eqnarray}
Where the blank positions in above matrices are zeroes.
Apparently if only the neutral (charged) components are picked out and the two Higgs fields are set to their vacuum expectation value, then up to some column or row switches and minus signs we will get the standard neutralino (chargino) mass mixing matrix.

When following the technique of adding $\ln(\slashed{p}+M+U)$ to $\ln(\slashed{p}-M-U)$ in~\cite{Henning:2014wua} to convert the above fermionic functional determinant to a bosonic-like one, for $U_L\neq U_R$ a further attention is needed. Note that with the identity $UU^\dag=U^\dag U$ under the implicit trace, it is $(\slashed{p}+U_LP_L+U_L^\dag P_R)(\slashed{p}-U_L^\dag P_L-U_LP_R)=p^2-(U_L^\dag U_L)(P_L+P_R)+\cdots=p^2-U_L^\dag U_L+\cdots$ (ignore $M$ which is always the same for both chiralities) that reproduces the canonical kinetic and mass terms, rather than $(\slashed{p}+U_LP_L+U_L^\dag P_R)(\slashed{p}-U_LP_L-U_L^\dag P_R)$. If we rewrite
\begin{equation}
U_LP_L+U_RP_R=U+U_5\gamma_5\qquad \text{with}\qquad U=\frac{U_L+ U_R}{2},\quad U_5= \frac{U_R-U_L}{2},\label{eq:UU5}
\end{equation}
it means to add $\ln(\slashed{p}+M+U-U_5\gamma_5)$ to $\ln(\slashed{p}-M-U-U_5\gamma_5)$, namely the $\slashed{p}\to-\slashed{p}$ in the trick in~\cite{Henning:2014wua} also induce a $\gamma_5\to-\gamma_5$. This should bring us no trouble because the trace containing odd number of $\gamma_5$ will give us CP-odd operaotrs and thus must vanish since we are considering the CP-even Lagragian in Eq. (\ref{eq:lagrangian}).

Finally, in the CDE, the Lagrangian reads\footnote{Note that we have the coefficient -1/4 instead of the usual -1/2 from the fermionic determinant (and after using the technique in ~\cite{Henning:2014wua}), because we have both the fermions and their conjugates in the $\chi$ in Eq.~(\ref{eq:S_eff0}).}
\begin{eqnarray}
\mathcal{L}_\text{eff} &=& -\frac{i}{4}\int\frac{d^4p}{(2\pi)^4}\text{tr}\ln\left[p^2-M^2-\widetilde{U}_{\text{ferm}}-\widetilde{G}\right],\label{eq:S_eff}
\label{eq:L_eff}
\end{eqnarray}
with
\begin{eqnarray}
U_{\text{ferm}}&=& \{M,~U\}+U^2-U_5^2-i[\slashed{D},U]-[M+U-i\slashed{D},~U_5]\gamma_5-\frac{g}{2}\sigma^{\mu \nu }F^a_{\mu \nu }t^a \\
\widetilde{U}_{\text{ferm}} &=& \sum_{n=0}^{\infty} \frac{(-i)^n}{n!}D_{\mu_1}\cdots D_{\mu_n}U_{\text{ferm}}	\frac{\partial^n}{\partial p_{\mu_1}\cdots\partial p_{\mu_n}} \, ,\\
\widetilde{F}_\mu&=&g\sum_{n=0}^{\infty} \frac{n+1}{(n+2)!}(-i)^{n+1}D_{\mu_1}D_{\mu_n}F^a_{\nu\mu}t^a\frac{\partial^{n+1}}{\partial p_{\mu_1}\cdots\partial p_{\mu_n}\partial p_\nu}\, ,\\
\widetilde{G}&=& -\{\widetilde{F}_\mu,~p^\mu\} -\widetilde{F}_\mu\widetilde{F}^\mu\, ,\label{eq:InputMatrix}
\end{eqnarray}
where $[~,~]$ and $\{~,~\}$ denote commutator and anti-commutator respectively,
and  $F_{\mu \nu}^a$ is the field strength of gauge field $A^a$.  The terms for the $U_5=0$ case have been explicitly expanded in~\cite{Huo:2015exa}.
Note that we adopt different regularization in dealing with the logarithm of Eq.~(\ref{eq:L_eff}) compared with~\cite{Huo:2015nka}, further relevant discussion is given in Appendix \ref{app:integrals}.

\section{Analytical Results and Check}
\label{sec:results}

\renewcommand\arraystretch{1.4}
\begin{table}[!h]
\centering
\caption{\label{dim6-operators} Dimension-6 SMEFT CP-even bosonic operators.}
\begin{tabular}{|rcl|rcl|}\hline
 \({\cal O}_{GG}\) &\(=\)& \(g_s^2 H^\dag H G_{\mu \nu }^aG^{a,\mu \nu }\) & \({\cal O}_H\)   &\(=\)& \(\frac{1}{2}\big(\partial_{\mu} H^\dag H\big)^2\)\\
 \({\cal O}_{WW}\) &\(=\)& \(g^2  H^\dag H W_{\mu \nu }^aW^{a,\mu \nu } \) &  \({\cal O}_T\)   &\(=\)& \(\frac{1}{2}\big(H^\dag \overleftrightarrow{D}_\mu H\big)^2\) \\
 \({\cal O}_{BB}\) &\(=\)& \(g'^2 H^\dag H B_{\mu \nu }B^{\mu \nu }\) & \({\cal O}_R\)   &\(=\)& \(H^\dag HD_{\mu}H^\dag D^\mu H\) \\
 \({\cal O}_{WB}\) &\(=\)& \(2gg'H^\dag {t^a}H W_{\mu \nu }^a B^{\mu \nu }\) &  \({\cal O}_D\)   &\(=\)& \(D^2H^\dag D^2H\) \\
 \({\cal O}_W\)   &\(=\)& \(ig(H^\dag\overleftrightarrow{D}_\mu t^a H )D_\nu W^{a\mu\nu}\) &  \({\cal O}_6\)   &\(=\)& \((H^\dag H)^3\) \\
 \({\cal O}_B\)   &\(=\)& \(ig'(H^\dag\overleftrightarrow{D}_\mu H)\partial_\nu B^{\mu\nu}\)  & \({\cal O}_{2G}\) &\(=\)& \(-\frac{1}{2} \big(D^\mu G_{\mu \nu }^a\big)^2\)\\
\({\cal O}_{3G}\) &\(=\)& \(\frac{1}{3!}g_sf^{abc}G_\rho ^{a\mu }G_\mu ^{b\nu }G_\nu ^{c\rho }\) &  \({\cal O}_{2W}\) &\(=\)& \(-\frac{1}{2} \big(D^\mu W_{\mu \nu }^a\big)^2\) \\
\({\cal O}_{3W}\) &\(=\)& \(\frac{1}{3!}g \epsilon^{abc}W_\rho ^{a\mu }W_\mu ^{b\nu }W_\nu ^{c\rho }\) & \({\cal O}_{2B}\) &\(=\)& \(-\frac{1}{2} \big(\partial^{\mu} B_{\mu \nu }\big)^2\) \\
 \({\cal O}_{HW}\)   &\(=\)& \(2ig(D_\mu H)^\dag t^a (D_\nu H)W^{a\mu\nu}\)  &  && \\
 \({\cal O}_{HB}\)   &\(=\)& \(2ig'Y_H(D_\mu H)^\dag (D_\nu H)B^{\mu\nu}\)  & & &\\
  \hline
\end{tabular}
\vspace{-10pt}
\end{table}

The target operators ${\cal O}_{i}$ are listed in Table.~\ref{dim6-operators}{\footnote{ This is a redundant basis, in the sense that ${\cal O}_{HW}$ and ${\cal O}_{HB}$ can be switched into $ {\cal O}_{WW}$, ${\cal O}_{BB}$, $ {\cal O}_{WB}$, $ {\cal O}_W$ and $ {\cal O}_B$, by using the relations:\begin{eqnarray}
{\cal O}_{HW} &=&{\cal O}_W-\frac{1}{4}({\cal O}_{WW}+{\cal O}_{WB}),\nonumber \\
{\cal O}_{HB} &=&{\cal O}_B-\frac{1}{4}({\cal O}_{BB}+{\cal O}_{WB}).
\end{eqnarray}
}}.
The analytical Wilson coefficients $c_i$  are shown in Appendix \ref{app:coefficients}.
Note that the coefficients of ${\cal O}_{2W}$, ${\cal O}_{3W}$, ${\cal O}_{2B}$ are obtained by applying the universal results of~\cite{Henning:2014wua} to the MSSM.
The unshown coefficients of ${\cal O}_{GG}$, ${\cal O}_{3G}$, ${\cal O}_{2G}$ are zeros. The coefficient of ${\cal O}_{6}$ is very tedious, and since it does not contribute to the currently interesting precision measurements we also omit it in this paper.

These results can be checked by inspecting the correction to $h\gamma\gamma$ effective coupling $\bar{c}_{h\gamma\gamma}$, which is defined as
\begin{equation}
\Delta\mathcal{L}_{h \gamma \gamma} = \bar{c}_{h \gamma \gamma} h A^{\mu \nu}A_{\mu \nu},
\label{eq:hgamma}
\end{equation}
where $A_{\mu\nu}=\partial_\mu A_\nu-\partial_\nu A_\mu$ is the electromagnetic field strength. On one hand, in the dimension-6 SMEFT, the $\bar{c}_{h\gamma\gamma}$ can be expressed as~\cite{Henning:2014wua}
\begin{equation}
\bar{c}_{h\gamma\gamma}=g^2\sin^2\theta_{w}v~(c_{WW}+c_{BB}-c_{WB}),
\end{equation}
where $v=246$ GeV is the Higgs vacuum expectation value. Plugging in the corresponding Wilson coefficients in Appendix \ref{app:coefficients}, we get
\begin{equation}
\bar{c}_{h \gamma \gamma} ^{EFT}= -\frac{g^4\sin^2\theta_{w}v\sin2\beta}{48\pi^2\mu M_2}.
\label{eq:caaL1}
\end{equation}
On the other hand, in the low energy theorem, the correction to the $h \gamma \gamma$ effective coupling from chargino loops can be calculated \cite{Ellis:1975ap,Shifman:1979eb,Carena:2012xa, Huo:2013fga}:
\begin{equation}
\bar{c}_{h \gamma \gamma}^{LET} = \frac{g^4\sin^2\theta_{w}v\sin2\beta}{48\pi^2\left(g^2v^2\sin2\beta-\mu M_2\right)}.
\label{eq:caaL2}
\end{equation}
From Eq.s (\ref{eq:caaL1}) and (\ref{eq:caaL2}) above one can see that the results derived from the two methods are consistent with each other when the value of $\mu M_2$ is much larger than that of $ m_w^2$.

\begin{figure}
\centering
\includegraphics[width=.45\textwidth]{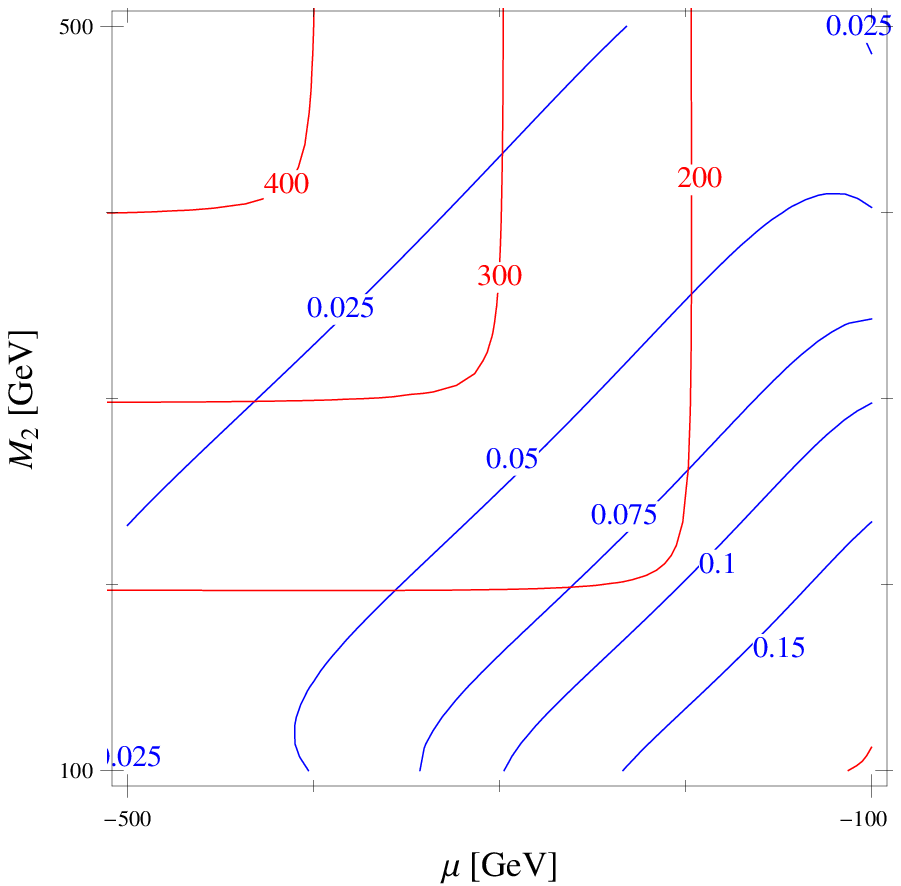}~
\includegraphics[width=.45\textwidth]{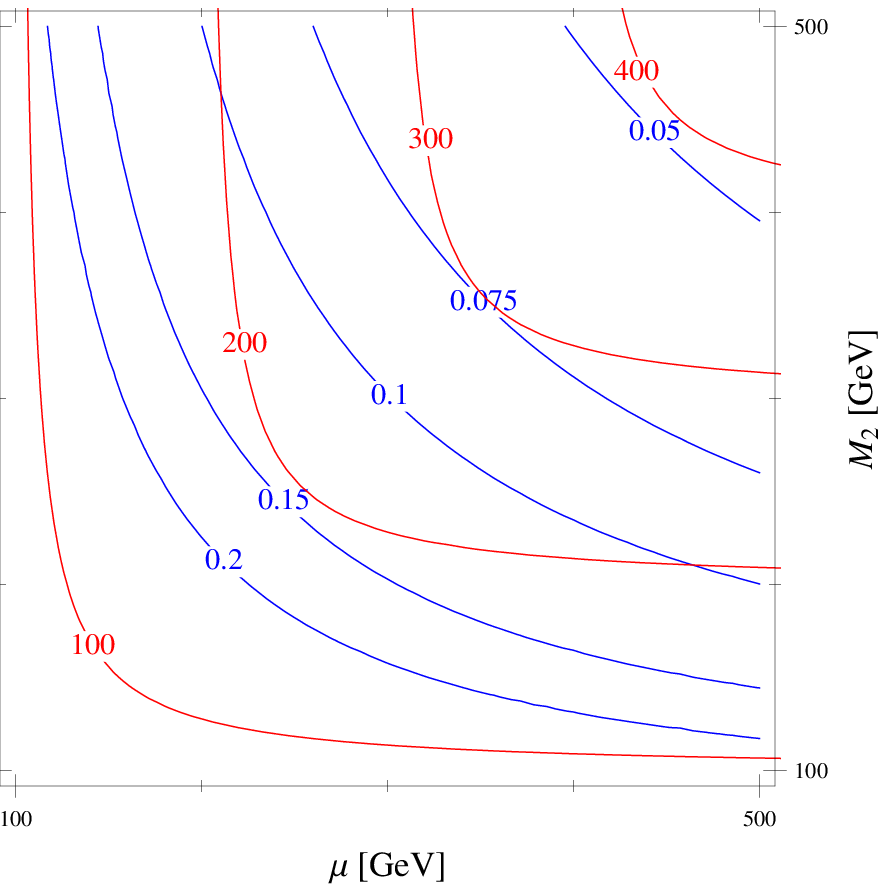}\\
\caption{Blue contours show the value of $\Delta R$ and red contours show the value of $m_{\tilde{\chi}_1^\pm}$. Here the results are independent of $M_1$ and we choose tan$\beta=2$.}
\label{fig:comp}
\end{figure}

To further illustrate the validity of the EFT we obtained, we also compare Eq.~(\ref{eq:caaL1}) with that from the  direct one-loop Feynman diagram calculation with charginos in the loop. In this loop calculation we use FeyncCalc \cite{Mertig:1990an, Shtabovenko:2016sxi} to calculate the amplitude with the help of the LoopTools \cite{Hahn:1998yk} to get the scalar loop integrals numerically.
In Fig.~\ref{fig:comp}, blue contours show the deviation of the two methods which is defined as $\Delta R=\frac{\bar{c}_{h \gamma \gamma} ^{EFT}}{\bar{c}_{h \gamma \gamma} ^{loop}}-1$~\cite{Drozd:2015kva}, and red curves show masses of the lighter chargino $m_{\tilde{\chi}_1^\pm}$ at tree level. One can see that the EFT works well when the charginos are heavier than about 200 GeV, characterized by $\Delta R <0.1$.

Moreover, we also calculate the oblique parameters $S$ and $T$ analytically to check our results. For simplicity, we assume that $\mu=M_1 = M_2=M$ and $\tan \beta = 1$, and we find that both the EFT results and the one-loop Feynman diagram  induced results are consistent. In this case, the $S$ and  $T$ are exactly the same within these two methods, i.e., 
\begin{eqnarray}
S &=& \frac{v^2\left({g'}^2 + 43 g^2 \right)}{240 \pi M^2}, \\
T &=& \frac{v^2\left({g'}^2 +  g^2 \right)}{40 \pi M^2}.
\end{eqnarray}

\section{Projected Future Constraints}
\label{sec:constraint}
In this section, we mainly discuss the above EFT contributions to the electroweak precision test (EWPT) (the oblique parameters $T$ and $S$), the triple gauge boson coupling (TGC) and Higgs production and decay rates, which will be measured with high precisions at the future lepton colliders such as the ILC, the CEPC and the FCC-ee.
The mapping of the Wilson coefficients onto these observables are well summarized in section 4 of Ref.~\cite{Henning:2014wua}.
Note that here we consider only the contribution from the electroweakino sector, rather than the whole MSSM which also includes the sfermions (see~\cite{Henning:2014gca,Huo:2015nka,Drozd:2015rsp}) and gluinos, as well as the extra Higgs (see, \emph{e.g.},~\cite{Chiang:2015ura}). For natural supersymmetry the stop sector usually has the largest contribution, but it is facing more and more severe experimental constraints.
In scenarios with heavy sfermions which are usually motivated by the neutralino lightest supersymmetric particle (LSP) dark matter candidate study, \emph{e.g.}, the split supersymmetry scenario, the dominant contribution could come from the electroweakino sector, and the collider search channel based on $\tilde{t}\to \tilde{\chi}^0_1+t$ becomes inaccessible. In this paper we assume such a scenario.

\begin{table}[th]
\caption{The uncertainties expected at each experiments, for the EWPT experiments ($T$ and $S$), the TGC experiments ($\Delta g_1^Z$, $\Delta\kappa_\gamma$ and $\lambda_\gamma$) and two representative channels of eight Higgs experiment channels (the others refer to the references therein).}
\label{tab:sigs}
\scriptsize{
\begin{tabular}{cccccccc}
\hline\hline
Observable & ~~$T$~~ & ~~$S$~~ & ~$10^4\Delta g_1^Z$~ & ~$10^4\Delta\kappa_\gamma$~ & ~$10^4\lambda_\gamma$~ & ~$\frac{\Delta(\sigma_{Zh}\text{Br}_{bb})}{\sigma_{Zh}^\text{SM}\text{Br}_{bb}^\text{SM}}$~ & ~$\frac{\Delta(\sigma_{Zh}\text{Br}_{\gamma\gamma})}{\sigma_{Zh}^\text{SM}\text{Br}_{\gamma\gamma}^\text{SM}}$~  \\
\hline
ILC (250~GeV) & $0.022$~\cite{Baak:2014ora} & $0.017$~\cite{Baak:2014ora} & & & & $1.1\%$~\cite{Baer:2013cma} & $35\%$~\cite{Baer:2013cma} \\
ILC (500~GeV) & $0.022$~\cite{Baak:2014ora} & $0.017$~\cite{Baak:2014ora} & $2.8$~\cite{AguilarSaavedra:2001rg} & $3.1$~\cite{AguilarSaavedra:2001rg} & $4.3$~\cite{AguilarSaavedra:2001rg} & $0.66\%$~\cite{Baer:2013cma} & $23\%$~\cite{Baer:2013cma} \\
ILC (1~TeV) & $0.022$~\cite{Baak:2014ora} & $0.017$~\cite{Baak:2014ora} & $1.8$~\cite{AguilarSaavedra:2001rg} & $1.9$~\cite{AguilarSaavedra:2001rg} & $2.6$~\cite{AguilarSaavedra:2001rg} & $0.47\%$~\cite{Baer:2013cma} & $8.5\%$~\cite{Baer:2013cma} \\
CEPC & $0.009$~\cite{Fan:2014vta} & $0.014$~\cite{Fan:2014vta} & $1.59$~\cite{Bian:2015zha} & $2.30$~\cite{Bian:2015zha} & $1.67$~\cite{Bian:2015zha} & $0.32\%$~\cite{Fan:2014vta} & $9.1\%$~\cite{Fan:2014vta} \\
FCC-ee & $0.004$~\cite{TeraZ} & $0.007$~\cite{TeraZ} & & & & $0.2\%$~\cite{Gomez-Ceballos:2013zzn} & $3.0\%$~\cite{Gomez-Ceballos:2013zzn} \\
\hline\hline
\end{tabular}}
\end{table}

The projected sensitivities of the proposed future lepton colliders are listed in Table~\ref{tab:sigs}. Naively we can see from the EWPT and Higgs data that the FCC-ee would be the best platform to perform such a precision test. So here we will combine the projected FCC-ee EWPT and Higgs measurements with the projected CEPC TGC measurements as our example of applying our EFT results to constrain model parameters.

\begin{figure}
\centering
\includegraphics[width=.45\textwidth]{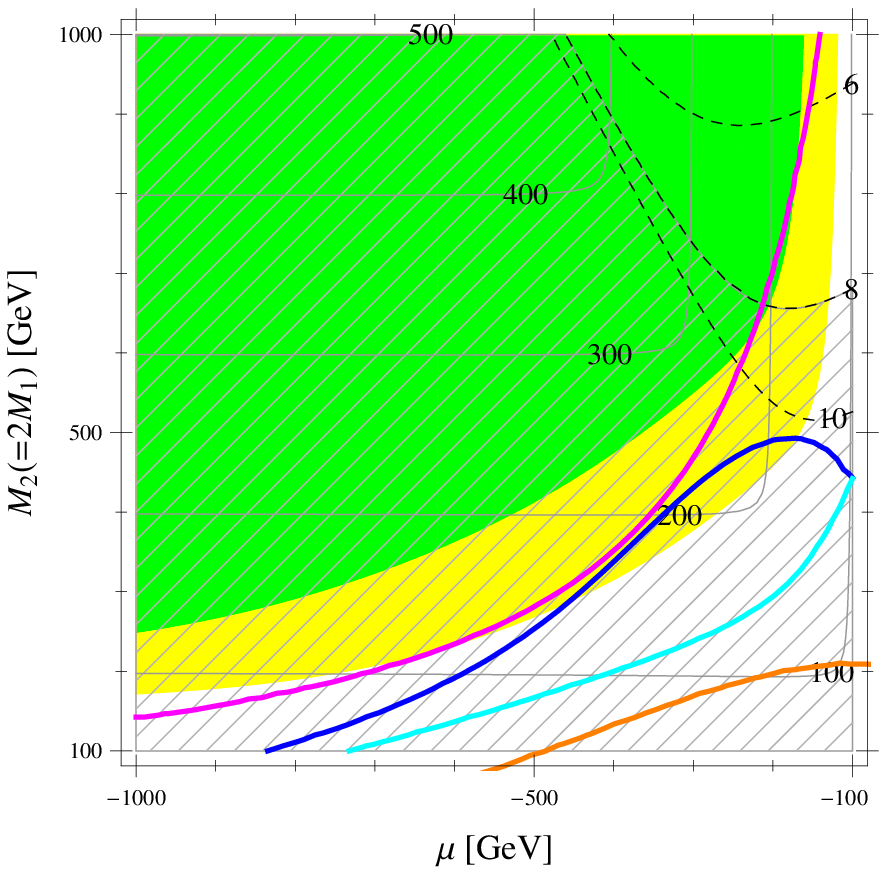}~
\includegraphics[width=.45\textwidth]{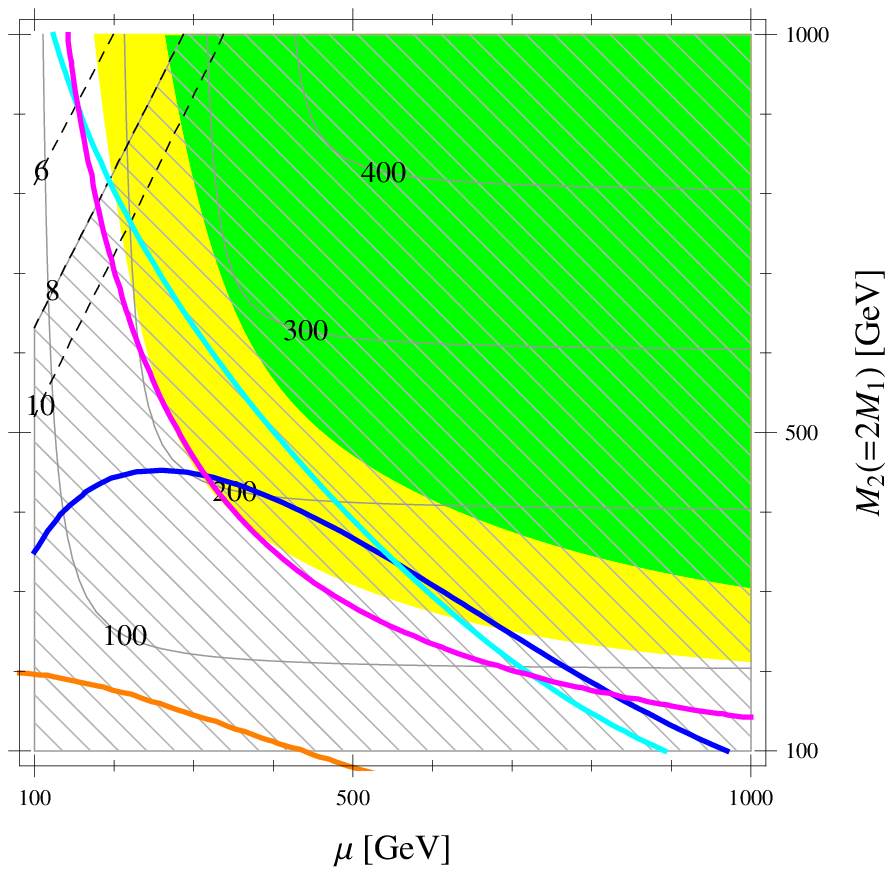}\\
\includegraphics[width=.45\textwidth]{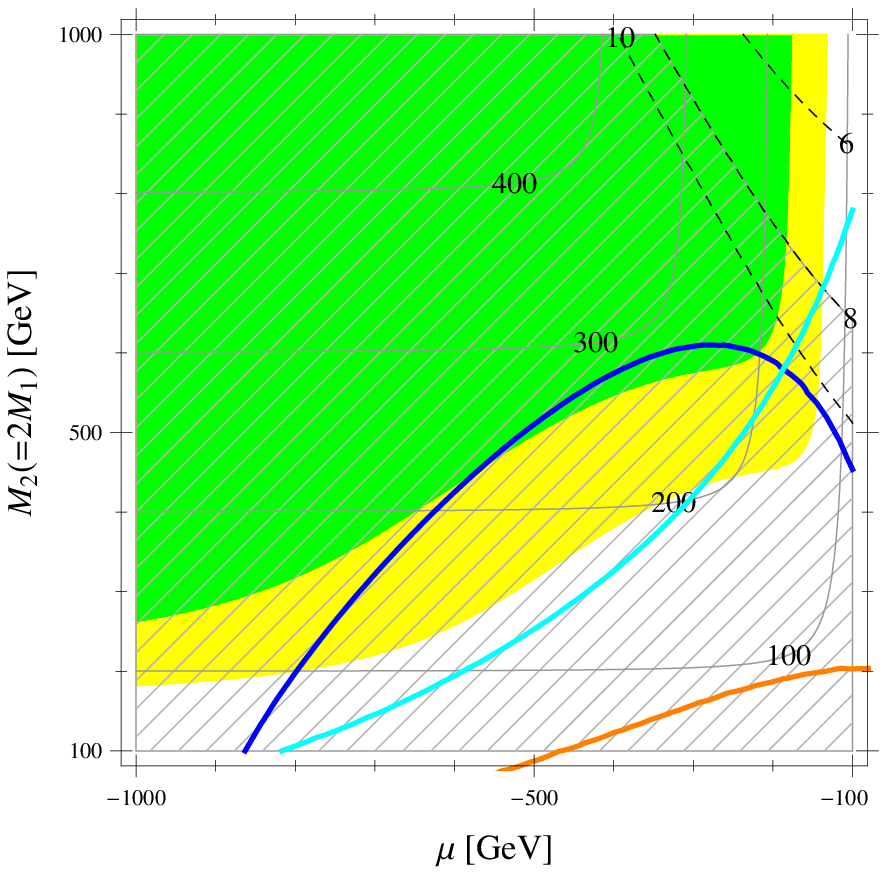}~
\includegraphics[width=.45\textwidth]{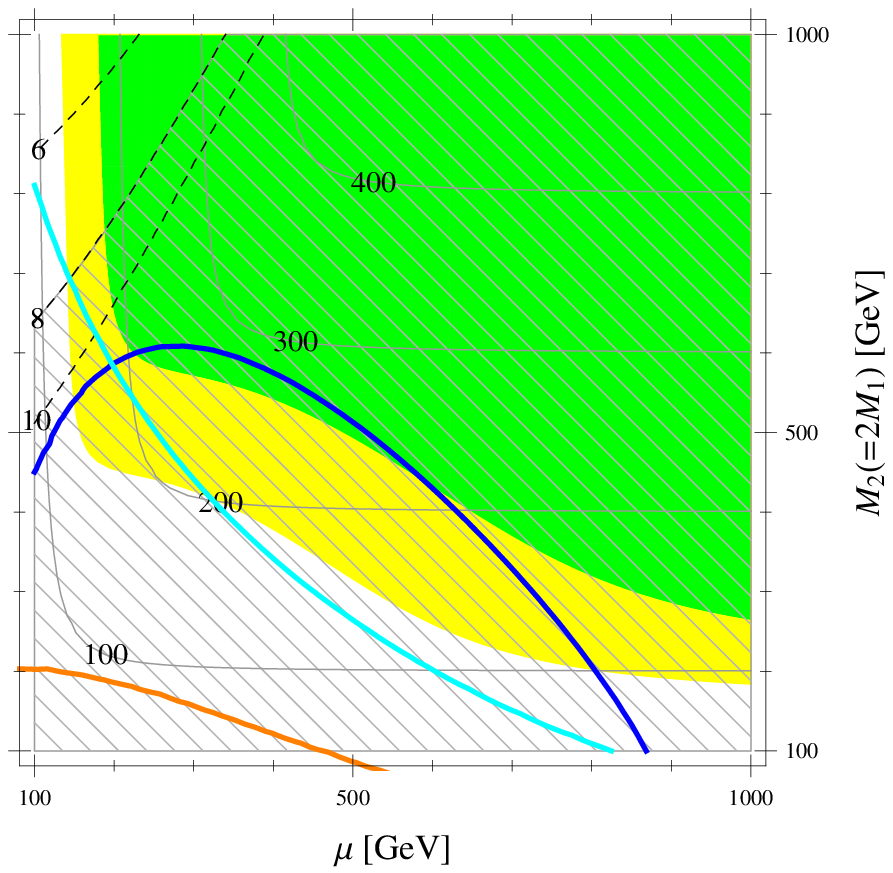}
\caption{The expected $\chi^2$ constraints from future CEPC and FCC-ee experiments for the gaugino unification scenario ($M_1/M_2=0.5$). The upper and lower panels stand for $\tan\beta=2$ and $\tan\beta=50$, respectively. The $1\sigma$ and $2\sigma$ allowed region are green and yellow shaded. The tree level neutralino LSP masses and the tree level mass splitting between the NLSP and LSP are shown as solid and dashed contours respectively, the $m_{\tilde{\chi}^0_2/\tilde{\chi}^\pm_1}-m_{\tilde{\chi}^0_1}>8$~GeV region which can also be detected by the soft lepton plus missing $E_T$ search is hatched. Also shown are several individual $1\sigma$ constraints (here covariance is not considered), including the FCC-ee EWPT T (blue) and S (cyan), the FCC-ee $h\to\gamma\gamma$ (magenta) and the CEPC $\Delta g_1^Z$ (orange), if they are strong enough to be shown.}
\label{fig:gaugino}
\end{figure}

\begin{figure}
\centering
\includegraphics[width=.45\textwidth]{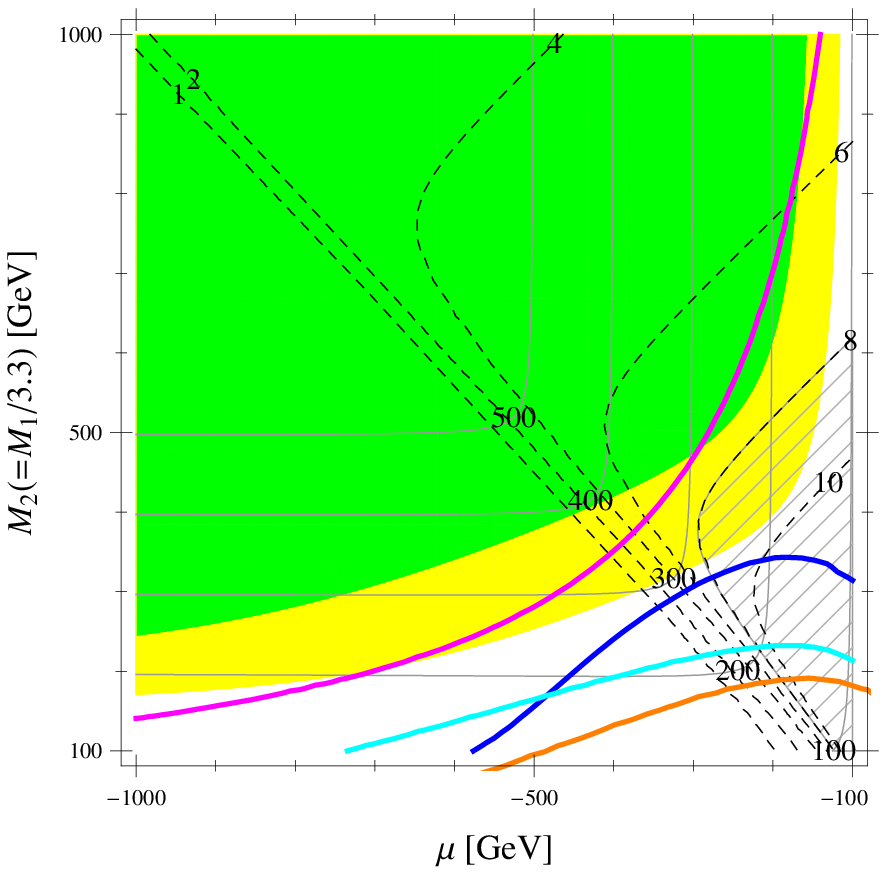}~
\includegraphics[width=.45\textwidth]{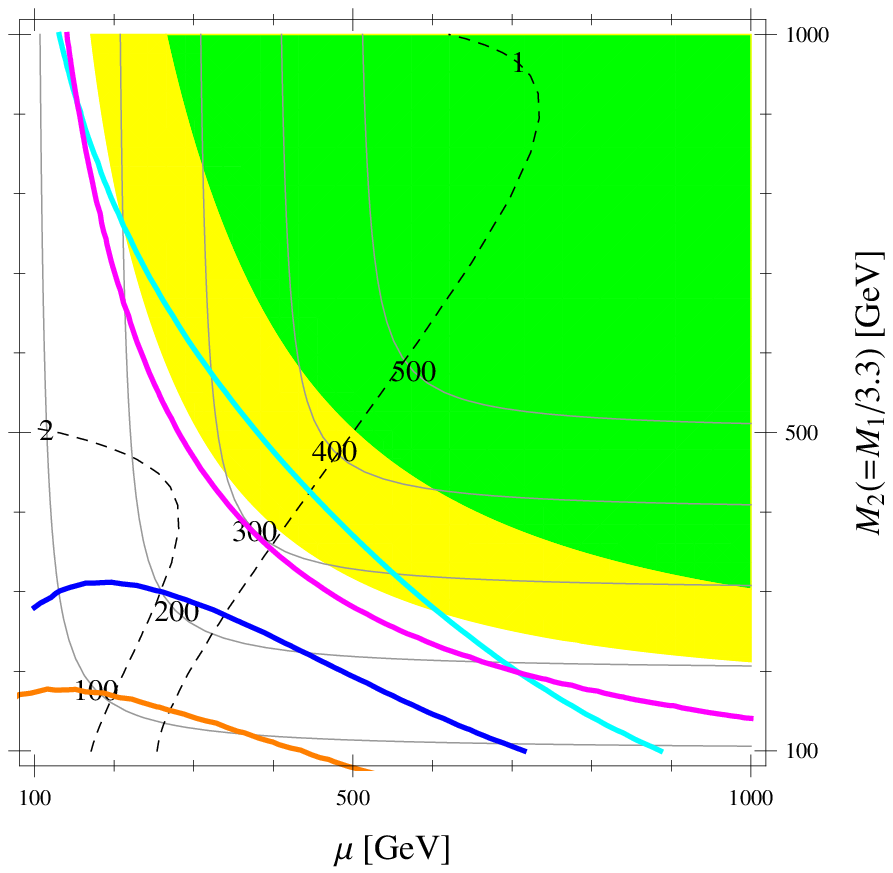}\\
\includegraphics[width=.45\textwidth]{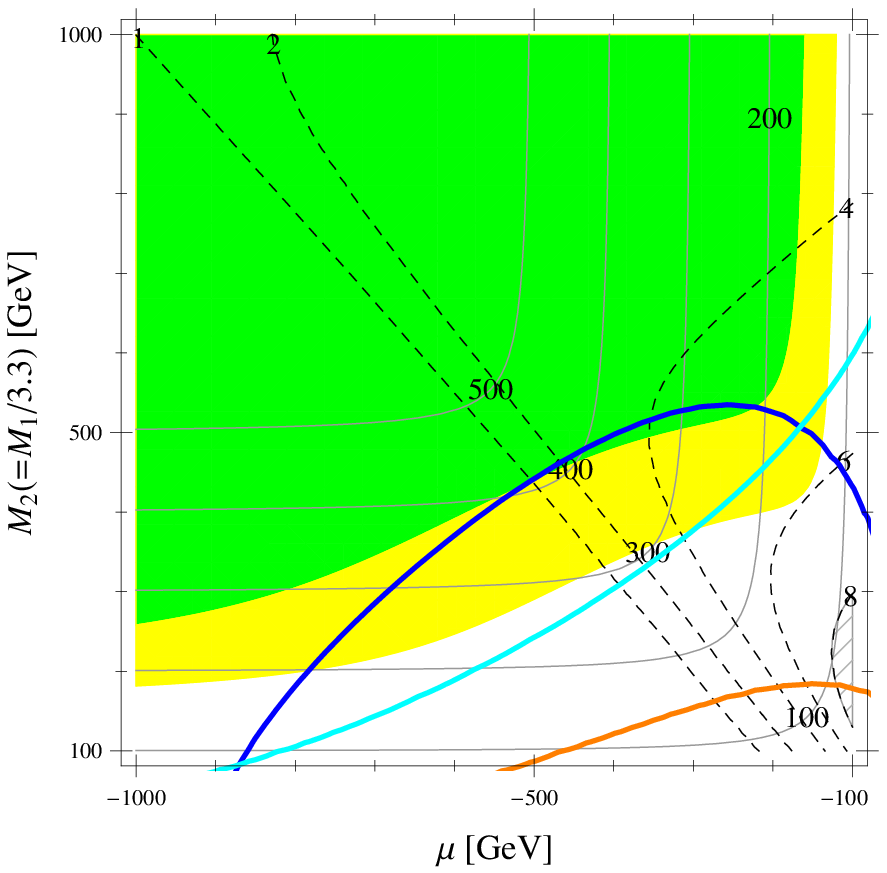}~
\includegraphics[width=.45\textwidth]{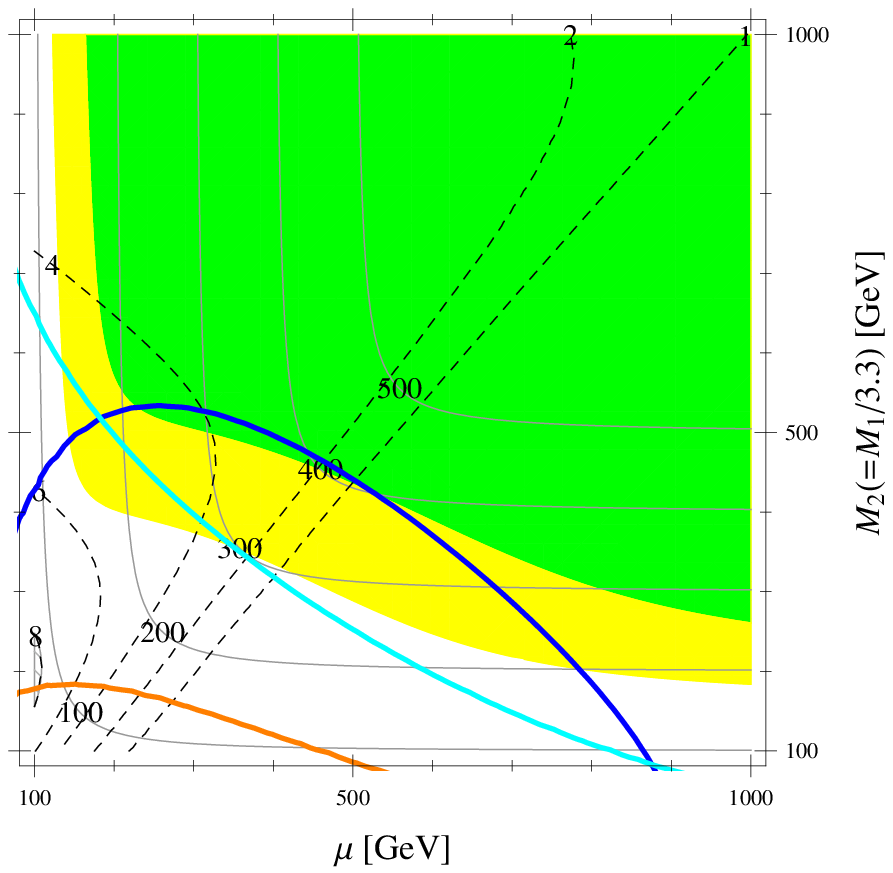}
\caption{The same as Fig.~\ref{fig:gaugino}, but for the anomaly mediation scenario ($M_1/M_2=3.3$).}
\label{fig:anomaly}
\end{figure}

The combined constraints from all three kinds of measurements, as the indirect searches for new physics,
are presented in Fig.~\ref{fig:gaugino} and Fig.~\ref{fig:anomaly}, for the gaugino mass unification scenario (Fig.~\ref{fig:gaugino}, $M_1=M_2/2$) and the anomaly mediation scenario (Fig.~\ref{fig:anomaly}, $M_1=3.3M_2$) respectively. In each scenario we choose a typical small $\tan\beta$ value of $2$ (upper panels) and a typical large $\tan\beta$ value of $50$ (lower panels), and show two choices of the sign of $\mu$ (left and right panels). The $1\sigma$ and $2\sigma$ allowed regions are green and yellow shaded. We can see that the $\chi^2$ constrained region on the $M_2$ vs.~$\mu$ plane are almost the same for the two different scenarios, which can be understood because the bino's contribution is not significant. The dependence on $\tan\beta$ is not strong on this plane either. And at large $\tan\beta$ the constraints are more symmetric between different $\mu$ choices.

The tree level neutralino LSP mass contours are shown as solid curves, and the tree level mass splitting between the neutralino LSP and the chargino or neutralino next-to-lightest supersymmetric particle (NLSP) contours are shown as dashed curves.
Unlike the shaded $\chi^2$ constrained region and the LSP mass contours, the mass splitting shows the greatest difference in different scenarios or $\tan\beta$ choices.
The significant degeneracy between  LSP and  NLSP only happens for almost pure higgsino LSP in the gaugino unification scenario, while in the anomaly mediation scenario the LSP is  degenerated with the NLSP in almost full parameter space.
Traditional direct collider electroweakino searches using the soft lepton plus missing $E_T$ channel become very inefficient for such degenerated electroweakino spectrum, \emph{e.g.}, the current CMS search is limited by $m_{\tilde{\chi}^\pm_1,\tilde{\chi}^2_0}-m_{\tilde{\chi}^1_0}\gtrsim8$~GeV~\cite{CMS:2017fij} (the meshed region in Fig.~\ref{fig:gaugino} and \ref{fig:anomaly}), and the future high luminosity LHC cannot significantly fill the gap. We can see that the indirect searches can be very useful in probing such a previously unbounded region, especially in the anomaly mediation scenario. Assuming null signal, the LSP can be constrained to be heavier than about $200$~GeV at $2\sigma$ level in the anomaly mediation scenario, regardless of the degeneracy of the spectrum.

In Figs.~\ref{fig:gaugino} and \ref{fig:anomaly}, also shown are the individual $1\sigma$ constraints from each best precision measurement in Table~\ref{tab:sigs}.
The ones strong enough to be visible include the FCC-ee EWPT constraints of $T$ (blue curve) and $S$ (cyan curve), the FCC-ee $h\to\gamma\gamma$ constraint (magenta curve) and the CEPC TGC $\Delta g_1^Z$ constraint (orange curve).
Because the EWPT $T$ and $S$ have some covariance and the $h\to\gamma\gamma$ branching ratio should be combined with the Higgs production cross section, the total constraint is not necessarily stronger than any single one of them. Since EWPT $T$ parameter is the only one proportional to the fourth power of the coupling of new physics to SM Higgs, in many models it gives the strongest constraint.
However here the SM $g\sim0.65$ is not large enough to give it an edge.
In the large $\tan\beta$ case $T$ is the most sensitive observable in the region $M_2\simeq\mu$, which dominates the central notch of the $\chi^2$ region.
The EWPT $S$ parameter is the most sensitive one in the $\mu>0$ and small $\tan\beta$ case, leading to the most stringent constraint on the $M_2$ vs.~$\mu$ plane. The $\tan\beta$ dependence of $h\to\gamma\gamma$ can be seen from Eq.~(\ref{eq:caaL1}), and at $\beta\sim\frac{\pi}{4} ~(\sin 2 \beta \sim 1)$ this channel can be significant,  especially if the Higgs total production cross section can be measured independently and precisely. The TGC $\Delta g_1^Z$ constraint alone looks not very impressive, but actually the covariance of the TGC experiments are in a direction such that the overall $\chi^2$ is still competitive compared with other experiments.

\begin{figure}
\centering
\includegraphics[width=.45\textwidth]{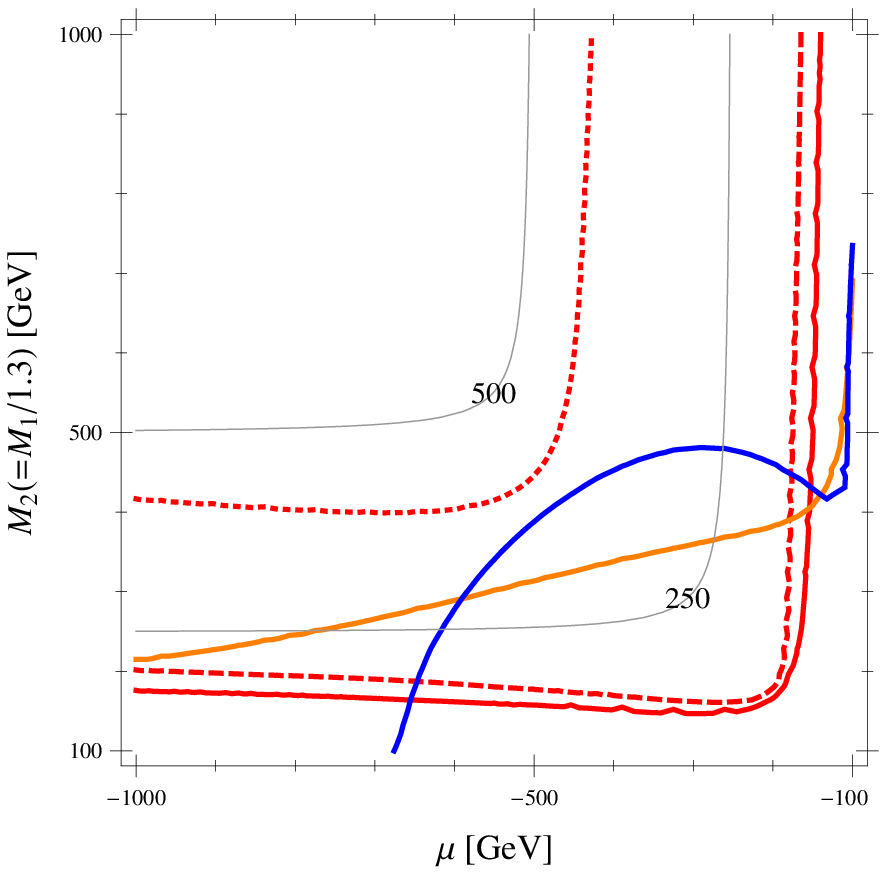}~
\includegraphics[width=.45\textwidth]{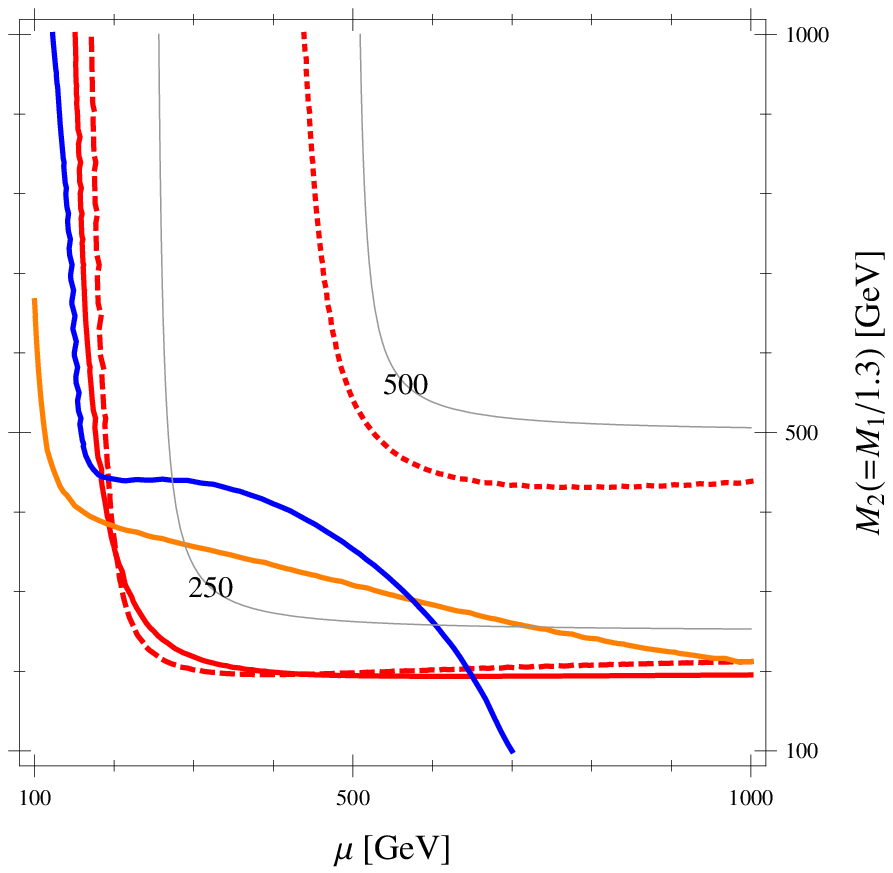}
\caption{The projected $1\sigma$ $\chi^2$ constraint of each of the best experiments, including FCC-ee EWPT experiments (blue curves), the CEPC TGC experiments (orange curves) and the FCC-ee Higgs precision measurements (red solid curves). Although the EFT approximation may break down, for illustrative purpose we also show ILC $500$~GeV and $1$~TeV run Higgs precision measurement constraints (red dashed curves and red dotted curves respectively). With the expected precisions reported in~\cite{Baer:2013cma} such sensitivities happen to be close but a little bit weaker than the naive direct production bounds ($m_\text{LSP}=\sqrt{s}/2$, grey curves). Here we choose $\tan\beta=10$ and $M_1/M_2=1.3$.}
\label{fig:amuse}
\end{figure}

In order to further illustrate our results, in particular with the effects of covariance of the EWPT and TGC experiments as well as the Higgs production cross section being imposed on the Higgs decay branching ratio measurements, in Fig.~\ref{fig:amuse} we show the $1\sigma$ allowed region from the projected FCC-ee EWPT experiments (blue curves), the projected CEPC TGC experiments (orange curves) and the projected FCC-ee Higgs precision measurements (red solid curves).
As is shown that $\tan\beta$ and $M_1/M_2$ is not important for the constraint on the $M_2$ vs.~$\mu$ plane (except for the NLSP LSP splitting), here we choose  general values, $\tan\beta=10$ and $M_1/M_2=1.3$.
Rather than the circular FCC-ee which has a fixed $\sqrt{s}$ of $240$ or $350$~GeV, the linear collider ILC can achieve a much higher $\sqrt{s}$, and in the EFT approximation scheme the Higgs production cross section correction may receive $\sqrt{s}$ boosted contributions (see Table 11 of~\cite{Henning:2014wua} and~\cite{Ellis:2017kfi}).
So we also show our EFT calculation of the projected ILC $500$~GeV and $1$~TeV run Higgs precision measurements (red dashed curves and red dotted curves respectively) as illustrations.
Although apparently the EFT approximation breaks down and the results are not reliable, and a consistent direct Feynman diagram calculation is beyond the scope of the current paper, we can still see the advantage of linear colliders.

\section{Summary and Conclusions}
\label{sec:conclude}

In this work, neutralinos and charginos in the MSSM are integrated out and the Wilson coefficients for the dimension-6 operators are presented analytically.  We have presented a convenient way of dealing two-component Weyl fermions in the covariant derivative expansion method. By adopting our results, we compute the $h\gamma\gamma$ effective coupling and oblique parameters, which are consistent with those in direct loop calculation method. And we believe that the SMEFT we obtained can be applied in more scenarios in a vast range of parameter space.
In the numerical evaluation, we use the projected electroweak oblique parameters $T$ and $S$, the TGC and  Higgs production and decay rates at future lepton collider to constrain the electroweakino sector.
While implementing the global fitting, we focus on the  gaugino unification scenario ($M_2\simeq2M_1$) and anomaly mediation scenario ($M_1\simeq3.3M_2$), and the allowed parameter space regions are obtained.
We also find that the constrained parameter space could stand out with higher center-of-mass collision energy combined with high experimental statistics.
In the parameter space where the neutralino as the lightest supersymmetric particle is very degenerated with the next-to-lightest  chargino/neutralino, the traditional collider search of electroweakino sector is inefficient. However, with the help of the Wilson coefficients of dimension-6 operators we obtained the future lepton collider like CEPC, ILC, FCC-ee  could make a difference in precision measurement experiments.

\acknowledgments{
H.H. is supported  by the China Postdoctoral Science Foundation under grand No. 2017M611008.
M.J. is supported by the NSFC under Contract No.~11775110,  No.~11690034  and No.~11405084
and  also supported by the European Union's Horizon 2020 Research and Innovation~(RISE)
programme under the Marie Sk\'lodowska-Curie grant agreement No.~644121,
and  the Priority Academic Program Development for Jiangsu Higher Education Institutions (PAPD).
J.S. is supported by the NSFC under grant No.11761141011, No.11747601, No.11690022 and No.11675243 and also supported by the Strategic Priority Research Program of the Chinese Academy of Sciences under grant No.XDB21010200 and No.XDB23030100.
}

\appendix

\section{Regularization Options}
\label{app:integrals}

In order to specify an order for the ${\textstyle\frac{\partial}{\partial p}}$ (contained in the $(\tilde{U}+\tilde{G})$) to act on the $(p^2-M^2)$, the CDE formulism is based on the following identity:
\begin{eqnarray}
(p^2-M^2-\tilde{U}-\tilde{G})^{-1}&=&(p^2-M^2)^{-1}+(p^2-M^2)^{-1}(\tilde{U}+\tilde{G})(p^2-M^2)^{-1} \nonumber \\
&&+(p^2-M^2)^{-1}(\tilde{U}+\tilde{G})(p^2-M^2)^{-1}(\tilde{U}+\tilde{G})(p^2-M^2)^{-1}\cdots.
\end{eqnarray}
However, the tricks used to get the $(p^2-M^2-\tilde{U}-\tilde{G})^{-1}$ from the $\ln(p^2-M^2-\tilde{U}-\tilde{G})$ have ambiguity. Two convenient options are
\begin{equation}
\ln(p^2-M^2-\tilde{U}-\tilde{G})=\int_1^\infty-d\xi\frac{d}{d\xi}\ln(p^2-\xi M^2-\tilde{U}-\tilde{G})=\int_1^\infty d\xi M^2\Big(p^2-\xi M^2-\tilde{U}-\tilde{G}\Big)^{-1}
\end{equation}
as in~\cite{Drozd:2015rsp,Cheyette:1985ue}, and
\begin{equation}
\ln(p^2-M^2-\tilde{U}-\tilde{G})=\int_0^\infty-du\frac{d}{du}\ln(p^2-M^2-\tilde{U}-\tilde{G}-u)=\int_0^\infty du\Big(p^2-M^2-\tilde{U}-\tilde{G}-u\Big)^{-1}
\end{equation}
as in~\cite{Huo:2015nka,Zuk:1985sw}. Both of the boundary terms at infinity ($\xi\to\infty$ and $u\to\infty$) are dropped implicitly. Correspondingly after finishing all the loop integrations, the implementation of the $\int_1^\infty d\xi$ in the first option and the $\int_0^\infty du$ in the second option, while making no difference in the degenerate case, clearly has a chance to make differences in the intermediate steps. For example, such an integration prototype of the expansion of effective action at the same order in above two options can be different,
\begin{align*}
\int_1^\infty d\xi\frac{M_1^2\text{ or }M_2^2}{\xi^{n+1}(xM_1^2+(1-x)M_2^2)^{n+1}}&=\frac{1}{n}\frac{M_1^2\text{ or }M_2^2}{(xM_1^2+(1-x)M_2^2)^{n+1}},\\
\int_0^\infty du\frac{1}{(xM_1^2+(1-x)M_2^2+u)^{n+1}}&=\frac{1}{n}\frac{1}{(xM_1^2+(1-x)M_2^2)^n}.
\end{align*}
However, it is checked that while the individual integration can differ with different regularization options, the whole results of all the integrations' summation will always agree with each other when including all the contributions correctly.

\section{Coefficients of Operators}
\label{app:coefficients}

Here we list the  Wilson coefficients $c_i$ of operators ${\cal{O}}_i$ in Table.~\ref{dim6-operators}, except the zero ones and that of $c_6$. We introduce the following shorthand notations:
\begin{equation}
\begin{array}{cccc}
r_1 = \frac{M_1}{\mu}, & & &  r_2 = \frac{M_2}{\mu}, \\
s_{2\beta} = {\rm sin} (2\beta), & & &   c_{4\beta} = {\rm cos} (4\beta). \\
\end{array}
\end{equation}

Then the Wilson coefficients are
\begin{eqnarray}
c_{2W} &=& \frac{g^2 \left(r_2^2+2\right)}{120 \pi ^2 \mu ^2 r_2^2}, \\
c_{2B} &=& \frac{g'^2}{120 \pi ^2 \mu ^2}, \\
c_{3W} &=& - \frac{g^2 \left(r_2^2+2\right)}{480 \pi ^2 \mu ^2 r_2^2}, \\
c_{WW} &=& g'^2 \bigg\{ -\frac{\left(2 \left(2 r_1^4+5 r_1^2-1\right) r_1 s_{2 \beta }+13 r_1^4-2 r_1^2+1\right)}{1536 \pi ^2 (r_1^2 -1)^3 \mu ^2}  \nonumber \\
&& \hspace{2em} + \frac{ \left(2 r_1 s_{2 \beta }+r_1^2+1\right) r_1^4 {\rm Log}[\frac{M_1^2}{\mu^2}]}{256 \pi ^2 (r_1^2 -1)^4 \mu ^2} \bigg\} \nonumber \\
&+& g^2 \bigg\{ -\frac{2 \left(6 r_2^6-25 r_2^4+5 r_2^2-16\right) s_{2 \beta }-17 r_2^5+10 r_2^3-53 r_2}{1536 \pi ^2 (r_2^2 - 1)^3 \mu ^2 r_2} \nonumber \\
&& \hspace{2em}-\frac{\left(r_2^4+4\right)  \left(2 r_2 s_{2 \beta }+r_2^2+1\right){\rm Log}[\frac{M_2^2}{\mu^2}]}{256 \pi ^2 (r_2^2 - 1)^4 \mu ^2} \bigg\}, \\
c_{BB} &=&  g'^2 \bigg\{ -\frac{ \left(2 \left(2 r_1^4+5 r_1^2-1\right) r_1 s_{2 \beta }+13 r_1^4-2 r_1^2+1\right)}{1536 \pi ^2 (r_1^2 - 1)^3 \mu ^2} \nonumber \\
&& \hspace{2em}
 + \frac{  r_1^4 \left(2 r_1 s_{2 \beta }+r_1^2+1\right){\rm Log}[\frac{M_1^2}{\mu^2}]}{256 \pi ^2 (r_1^2 - 1)^4 \mu ^2} \bigg\} \nonumber \\
 &+& g^2  \bigg\{-\frac{2 \left(2 r_2^4+5 r_2^2-1\right) r_2 s_{2 \beta }+13 r_2^4-2 r_2^2+1}{512 \pi ^2 (r_2^2 - 1)^3 \mu ^2} \nonumber \\
&& \hspace{2em}
 + \frac{3 r_2^4  \left(2 r_2 s_{2 \beta }+r_2^2+1\right) {\rm Log}[\frac{M_2^2}{\mu^2}]}{256 \pi ^2 (r_2^2 - 1)^4 \mu ^2} \bigg\},\\
 c_{WB}&=&  g'^2 \bigg\{-\frac{ \left(2 \left(2 r_1^4+5 r_1^2-1\right) r_1 s_{2 \beta }+13 r_1^4-2 r_1^2+1\right)}{768 \pi ^2 (r_1^2 - 1)^3 \mu ^2}  \nonumber \\
&& \hspace{2em}   +  \frac{r_1^4   \left(2 r_1 s_{2 \beta }+r_1^2+1\right) {\rm Log}[\frac{M_1^2}{\mu^2}]}{128 \pi ^2 (r_1^2 - 1)^4 \mu ^2}
\bigg\} \nonumber \\
 &+& g^2 \bigg\{ \frac{\left(4 r_2^5-38 r_2^3+46 r_2\right) s_{2 \beta }-11 r_2^4-2 r_2^2+25}{768 \pi ^2 (r_2^2 - 1)^3 \mu ^2} \nonumber \\
&& \hspace{2em}
 +\frac{\left(r_2^4-2\right) {\rm Log}[\frac{M_2^2}{\mu^2}] \left(2 r_2 s_{2 \beta }+r_2^2+1\right)}{128 \pi ^2 (r_2^2 - 1)^4 \mu ^2}
\bigg\},  \\
 c_{W}&=&  g'^2 \bigg\{ \frac{4 \left(7 r_1^4-11 r_1^2-2\right) r_1 s_{2 \beta }+23 r_1^6-17 r_1^4-35 r_1^2+5}{1152 \pi ^2 (r_1^2 - 1)^4 \mu ^2} \nonumber \\
&& \hspace{2em}
 -\frac{r_1^3  \left(2 \left(r_1^4-3\right) s_{2 \beta }+\left(r_1^4+4 r_1^2-9\right) r_1\right){\rm Log}[\frac{M_1^2}{\mu^2}]}{192 \pi ^2 (r_1^2 - 1)^5 \mu ^2}
\bigg\} \nonumber \\
 &+& g^2 \bigg\{ -\frac{4 \left(5 r_2^4+11 r_2^2-10\right) r_2 s_{2 \beta }+r_2^6+41 r_2^4+11 r_2^2-29}{384 \pi ^2 (r_2^2 - 1)^4 \mu ^2}  \nonumber \\
&&
+ \frac{ \left(2 \left(r_2^6+12 r_2^4-3 r_2^2-4\right) r_2 s_{2 \beta }+r_2^8+4 r_2^6+27 r_2^4-16 r_2^2-4\right){\rm Log}[\frac{M_2^2}{\mu^2}]}{192 \pi ^2 (r_2^2 - 1)^5 \mu ^2}
\bigg\},  \\
 c_{B}&=&  g'^2 \bigg\{ \frac{4 \left(7 r_1^4-11 r_1^2-2\right) r_1 s_{2 \beta }+23 r_1^6-17 r_1^4-35 r_1^2+5}{1152 \pi ^2 (r_1^2 - 1)^4 \mu ^2}  \nonumber \\
&& \hspace{2em}
 -\frac{r_1^3  \left(2 \left(r_1^4-3\right) s_{2 \beta }+\left(r_1^4+4 r_1^2-9\right) r_1\right){\rm Log}[\frac{M_1^2}{\mu^2}]}{192 \pi ^2 (r_1^2 - 1)^5 \mu ^2}
\bigg\} \nonumber \\
 &+& g^2 \bigg\{ \frac{4 \left(7 r_2^4-11 r_2^2-2\right) r_2 s_{2 \beta }+23 r_2^6-17 r_2^4-35 r_2^2+5}{384 \pi ^2 (r_2^2 - 1)^4 \mu ^2}   \nonumber \\
&& \hspace{2em}  -\frac{r_2^3  \left(2 \left(r_2^4-3\right) s_{2 \beta }+\left(r_2^4+4 r_2^2-9\right) r_2\right){\rm Log}[\frac{M_2^2}{\mu^2}]}{64 \pi ^2 (r_2^2 - 1)^5 \mu ^2}
\bigg\},  \\
 c_{HW}&=&  g'^2 \bigg\{ -\frac{(r_1^2 - 1)-2 \left(r_1^5+10 r_1^3+r_1\right) s_{2 \beta }+11 r_1^6-35 r_1^4}{384 \pi ^2 (r_1^2 - 1)^4 \mu ^2}   \nonumber \\
&& \hspace{2em}  + \frac{r_1^3 \left(r_1^2+1\right) \left(r_1^3-3 r_1-2 s_{2 \beta }\right)  {\rm Log}[\frac{M_1^2}{\mu^2}]}{64 \pi ^2 (r_1^2 - 1)^5 \mu ^2}
\bigg\} \nonumber \\
 &+& g^2 \bigg\{ \frac{2 \left(r_2^5+10 r_2^3+r_2\right) s_{2 \beta }+5 r_2^6-13 r_2^4+47 r_2^2-15}{128 \pi ^2 (r_2^2 - 1)^4 \mu ^2}   \nonumber \\
&& \hspace{2em}  -\frac{\left(r_2^2+1\right)  \left(6 r_2^3 s_{2 \beta }+r_2^6-3 r_2^4+12 r_2^2-4\right){\rm Log}[\frac{M_2^2}{\mu^2}]}{64 \pi ^2 (r_2^2 - 1)^5 \mu ^2}
\bigg\},  \\
 c_{HB}&=&  g'^2 \bigg\{ -\frac{(r_1^2 - 1)-2 \left(r_1^5+10 r_1^3+r_1\right) s_{2 \beta }+11 r_1^6-35 r_1^4}{384 \pi ^2 (r_1^2 - 1)^4 \mu ^2}  \nonumber \\
&& \hspace{2em}  +\frac{r_1^3 \left(r_1^2+1\right) \left(r_1^3-3 r_1-2 s_{2 \beta }\right) {\rm Log}[\frac{M_1^2}{\mu^2}]}{64 \pi ^2 (r_1^2 - 1)^5 \mu ^2}
\bigg\} \nonumber \\
 &+& g^2 \bigg\{ \frac{2 \left(r_2^5+10 r_2^3+r_2\right) s_{2 \beta }+5 r_2^6-13 r_2^4+47 r_2^2-15}{128 \pi ^2 (r_2^2 - 1)^4 \mu ^2}   \nonumber \\
&& \hspace{2em}  -\frac{\left(r_2^2+1\right)  \left(6 r_2^3 s_{2 \beta }+r_2^6-3 r_2^4+12 r_2^2-4\right){\rm Log}[\frac{M_2^2}{\mu^2}]}{64 \pi ^2 (r_2^2 - 1)^5 \mu ^2}
\bigg\},  \hspace{3em}\\
 c_{D}&=&  g'^2 \bigg\{ \frac{-\left(r_1^5+10 r_1^3+r_1\right) s_{2 \beta }+r_1^6-7 r_1^4-7 r_1^2+1}{96 \pi ^2 (r_1^2 - 1)^4 \mu ^2}   \nonumber \\
&& \hspace{2em}  +\frac{r_1^3  \left(\left(r_1^2+1\right) s_{2 \beta }+2 r_1\right){\rm Log}[\frac{M_1^2}{\mu^2}]}{16 \pi ^2 (r_1^2 - 1)^5 \mu ^2}
\bigg\} \nonumber \\
 &+& g^2 \bigg\{ \frac{-\left(r_2^5+10 r_2^3+r_2\right) s_{2 \beta }+r_2^6-7 r_2^4-7 r_2^2+1}{32 \pi ^2 (r_2^2 - 1)^4 \mu ^2}   \nonumber \\
&& \hspace{2em}  +\frac{3 r_2^3  \left(\left(r_2^2+1\right) s_{2 \beta }+2 r_2\right){\rm Log}[\frac{M_2^2}{\mu^2}]}{16 \pi ^2 (r_2^2 - 1)^5 \mu ^2}\bigg\},
\end{eqnarray}
\begin{eqnarray}
c_H &=& g'^4 \bigg\{  \frac{1}{1536 \pi ^2 (r_1^2 - 1)^4 \mu ^2} \Big[ (2 r_1^6-45 r_1^4-126 r_1^2+25) c_{4 \beta }   \nonumber \\
 && \hspace{2em} + 12 \left(r_1^4+52 r_1^2-5\right)r_1 s_{2 \beta } -28 r_1^6+309 r_1^4+192 r_1^2-41 \Big]
\nonumber \\
&&+ \frac{   {\rm Log}[\frac{M_1^2}{\mu^2}] }{256 \pi ^2 (r_1^2 - 1)^5 \mu ^2} \Big[ (r_1^6+19 r_1^4+6 r_1^2-2) c_{4 \beta }   \nonumber \\
 && \hspace{2em}  +4 \left(r_1^6-15 r_1^4-12 r_1^2+2\right) r_1 s_{2 \beta }+2 r_1^8-15 r_1^6-59 r_1^4-2 r_1^2+2   \Big]  \nonumber \\
&+& g'^2 g^2 \bigg\{\frac{ {\rm Log}[\frac{M_1^2}{\mu^2}] r_1^3}{256 \pi ^2 (r_1^2 - 1)^4 (r_1 -r_2)(r_1 + r_2)\mu ^2} \Big[ -8 (r_1 + r_2)(r_1^2 -5) r_1 s_{2 \beta }  \nonumber \\
 && \hspace{2em} + (r_1^5+2 r_2 r_1^4-4 r_1^3-2 r_2 r_1^2-7 r_1-6 r_2) c_{4 \beta }  -7 r_1^5-6 r_2 r_1^4 \nonumber \\
 && \hspace{2em} +20 r_1^3+22 r_2 r_1^2+9 r_1+10 r_2 \Big] \nonumber \\
&+& \frac{ {\rm Log}[\frac{M_2^2}{\mu^2}]r_2^3 }{256 \pi ^2 (r_2^2 - 1)^4 (r_1 -r_2)(r_1 + r_2)\mu ^2} \Big[ 8 (r_1 + r_2)(r_2^2 -5 ) r_2 s_{2 \beta }  \nonumber \\
 && \hspace{2em} + (r_1^5+2 r_2 r_1^4-4 r_1^3-2 r_2 r_1^2-7 r_1-6 r_2)  c_{4 \beta }  \nonumber \\
 && \hspace{2em}   + 7 r_2^5+6 r_1 r_2^4-20 r_2^3-22 r_1 r_2^2-9 r_2-10 r_1 \Big] \nonumber \\
&+& \frac{1}{768 \pi ^2 (r_1^2 - 1)^3 (r_2^2 - 1)^3 \mu ^2}   \Big[ c_{4 \beta } \Big( 3 \left(4 r_2^4-9 r_2^2-1\right) r_2 r_1^5  \nonumber \\
 && \hspace{2em} +\left(-16 r_2^4-19 r_2^2+5\right) r_1^4 +3 \left(-9 r_2^4+14 r_2^2+7\right) r_2 r_1^3 \nonumber \\
 && \hspace{2em} +\left(-19 r_2^4+110 r_2^2-31\right) r_1^2 -3 r_2 \left(r_2^4-7 r_2^2+12\right) r_1+5 r_2^4-31 r_2^2-4\Big)   \nonumber \\
 && \hspace{2em} - 4 s_{2 \beta }  \Big(\left(r_2^4-32 r_2^2+7\right) r_1^5+\left(r_2^4-32 r_2^2+7\right) r_2 r_1^4 \nonumber \\
 && \hspace{2em} -4 \left(8 r_2^4-25 r_2^2+5\right) r_1^3-4 r_2 \left(8 r_2^4-25 r_2^2+5\right) r_1^2 \nonumber \\
 && \hspace{2em} +\left(7 r_2^4-20 r_2^2-11\right) r_1+\left(7 r_2^4-20 r_2^2-11\right) r_2\Big)   \nonumber \\
 && \hspace{2em} + \left(-16 r_2^5+101 r_2^3-7 r_2\right) r_1^5+\left(28 r_2^4 +61 r_2^2-23\right) r_1^4 \nonumber \\
 && \hspace{2em} + \left(101 r_2^4-262 r_2^2+5\right) r_2 r_1^3+\left(61 r_2^4-290 r_2^2+97\right) r_1^2  \nonumber \\
 && \hspace{2em} +\left(-7 r_2^5+5 r_2^3+80 r_2\right) r_1-23 r_2^4+97 r_2^2-8 \Big] \bigg\} \nonumber \\
&+& g^4 \bigg\{ \frac{1}{1536 \pi ^2 (r_2^2 - 1)^4 \mu ^2} \Big[  (30 r_2^6-169 r_2^4-250 r_2^2+53) c_{4 \beta }  \nonumber \\
 && \hspace{2em} -4 \left(29 r_2^4-412 r_2^2+47\right) r_2 s_{2 \beta }-112 r_2^6+681 r_2^4+564 r_2^2-125 \Big] \nonumber \\
&-& \frac{{\rm Log}[\frac{M_2^2}{\mu^2}]}{256 \pi ^2 (r_2^2 - 1)^5 \mu ^2} \Big[ (4 r_2^8-13 r_2^6-35 r_2^4-18 r_2^2+6) c_{4 \beta } \nonumber \\
 && \hspace{2em} -4 \left(5 r_2^6-35 r_2^4-32 r_2^2+6\right) r_2 s_{2 \beta }-6 r_2^8+19 r_2^6+155 r_2^4+6 r_2^2-6 \Big]
\bigg\}, \\
c_R &=& g'^4 \bigg\{ \frac{1}{384 \pi ^2 (r_1^2 - 1)^4 \mu ^2} \Big[ \left(r_1^6-10 r_1^4-37 r_1^2-2\right) c_{4 \beta }  \nonumber \\
 && \hspace{2em} +2 \left(-7 r_1^4+92 r_1^2+11\right) r_1 s_{2 \beta }-18 r_1^6+87 r_1^4+78 r_1^2-3 \Big] \nonumber \\
&+&  \frac{   {\rm Log}[\frac{M_1^2}{\mu^2}] r_1^2 } {64 \pi ^2 (r_1^2 - 1)^5 \mu ^2} \Big[ \left(5 r_1^2+3\right) c_{4 \beta }  \nonumber \\
 && \hspace{2em}  +2 \left(r_1^4-7 r_1^2-10\right) r_1 s_{2 \beta }+r_1^6-2 r_1^4-20 r_1^2-3 \Big]
\bigg\} \nonumber \\
&+& g'^2 g^2 \bigg\{  \frac{-r_1^3 {\rm Log}[\frac{M_1^2}{\mu^2}]}{128 \pi ^2 (r_1^2 - 1)^4 (r_1 - r_2)^3 (r_1 + r_2)\mu ^2} \Big[  -8 s_{2 \beta } \Big( r_1^6+r_2 r_1^5   \nonumber \\
 && \hspace{2em}  +3 r_1^4-5 r_2 r_1^3 -\left(5 r_2^2+1\right) r_1^2 +\left(3 r_2^3+r_2\right) r_1 + 2 r_2^2\Big)  \nonumber \\
 && \hspace{2em}  + c_{4 \beta } \Big(r_1^7-\left(3 r_2^2+2\right) r_1^5+2 \left(r_2^2+6\right) r_2 r_1^4  +\left(9-6 r_2^2\right) r_1^3   \nonumber \\
 && \hspace{2em} -20 r_2 r_1^2-\left(3 r_2^2+4\right) r_1+6 r_2^3+8 r_2\Big) -r_1^7-4 r_2 r_1^6-5 r_2^2 r_1^5  \nonumber \\
 && \hspace{2em} -18 r_1^5+2 r_2^3 r_1^4+16 r_2 r_1^4+26 r_2^2 r_1^3-5 r_1^3-12 r_2^3 r_1^2+20 r_2 r_1^2  \nonumber \\
 && \hspace{2em} -9 r_2^2 r_1+4 r_1-6 r_2^3-8 r_2 \Big] \nonumber \\
&+& \frac{r_2^3 {\rm Log}[\frac{M_2^2}{\mu^2}]}{128 \pi ^2 (r_2^2 - 1)^4 (r_1 - r_2)^3 (r_1 + r_2)\mu ^2} \Big[ -8 \Big((3 r_2 r_1^3+\left(2-5 r_2^2\right) r_1^2  \nonumber \\
 && \hspace{2em} +\left(r_2^5-5 r_2^3+r_2\right) r_1   +\left(r_2^4+3 r_2^2-1\right) r_2^2\Big) s_{2 \beta }  \nonumber \\
 && \hspace{2em} + \Big(2 \left(r_2^4+3\right) r_1^3-3 r_2 \left(r_2^2+1\right){}^2 r_1^2 \nonumber \\
 && \hspace{2em} +4 \left(3 r_2^4-5 r_2^2+2\right) r_1+\left(r_2^6-2 r_2^4+9 r_2^2-4\right) r_2\Big) c_{4 \beta } \nonumber \\
 && \hspace{2em}  -r_2^7-4 r_1 r_2^6-5 r_1^2 r_2^5-18 r_2^5+2 r_1^3 r_2^4+16 r_1 r_2^4+26 r_1^2 r_2^3  \nonumber \\
 && \hspace{2em} -5 r_2^3-12 r_1^3 r_2^2+20 r_1 r_2^2-9 r_1^2 r_2+4 r_2-6 r_1^3-8 r_1   \Big] \nonumber \\
&-& \frac{1}{384 \pi ^2 (r_1^2 - 1)^3 (r_2^2 - 1)^3 (r_1 - r_2)^2 \mu ^2}\Big[ 8 \left(r_2^6-11 r_2^4+16 r_2^2-3\right) r_1^3 r_2   \nonumber \\
 && \hspace{2em}  +2 \left(4 r_2^6+3 r_2^4-12 r_2^2+5\right) r_1 r_2 +c_{4 \beta } \Big( -\left(16 r_2^6+27 r_2^4-12 r_2^2+5\right) r_1^6  \nonumber \\
 && \hspace{2em}  + 8 \left(r_2^5+r_2^3+r_2\right) r_1^7 +\left(8 r_2^7+26 r_2^5-88 r_2^3+6 r_2\right) r_1^5-5 r_2^6+28 r_2^4-11 r_2^2  \nonumber \\
 && \hspace{2em}  +\left(-27 r_2^6+184 r_2^4-101 r_2^2+28\right) r_1^4+\left(12 r_2^6-101 r_2^4+40 r_2^2-11\right) r_1^2 \Big)  \nonumber \\
 && \hspace{2em}  -12 s_{2 \beta } \Big( \left(2 r_2^4+5 r_2^2-1\right) r_1^7+\left(-4 r_2^5-5 r_2^3+3 r_2\right) r_1^6+\left(-4 r_2^6+8 r_2^4-31 r_2^2+9\right) r_1^5   \nonumber \\
 && \hspace{2em}  +\left(2 r_2^6+8 r_2^4+17 r_2^2-9\right) r_2 r_1^4  + 3 \left(r_2^2-3\right) r_1 r_2^4+\left(5 r_2^4-31 r_2^2+8\right) r_1^2 r_2^3 \nonumber \\
 && \hspace{2em} -\left(r_2^4-9 r_2^2+2\right) r_2^3+\left(-5 r_2^6+17 r_2^4+8 r_2^2-2\right) r_1^3 \Big) \nonumber \\
 && \hspace{2em} +2 \left(r_2^5-26 r_2^3+r_2\right) r_1^7+\left(20 r_2^6+9 r_2^4+24 r_2^2-17\right) r_1^6+2 \left(r_2^6+r_2^4+97 r_2^2-15\right) r_2 r_1^5 \nonumber \\
 && \hspace{2em} + \left(9 r_2^6-344 r_2^4+271 r_2^2-68\right) r_1^4 -17 r_2^6-68 r_2^4+25 r_2^2  \nonumber \\
 && \hspace{2em} -2 \left(26 r_2^6-97 r_2^4+236 r_2^2-69\right) r_1^3 r_2+2 \left(r_2^6-15 r_2^4+69 r_2^2-19\right) r_1 r_2 \nonumber \\
 && \hspace{2em} +\left(24 r_2^6+271 r_2^4-164 r_2^2+25\right) r_1^2  \Big] \bigg\} \nonumber \\
&+& g^4 \bigg\{ -\frac{1}{384 \pi ^2 (r_2^2 - 1)^4 \mu ^2} \Big[ \left(5 r_2^6+76 r_2^4+103 r_2^2+8\right) c_{4 \beta }   \nonumber \\
 && \hspace{2em}  -2 \left(17 r_2^4+332 r_2^2+35\right) r_2 s_{2 \beta } +36 r_2^6-321 r_2^4-312 r_2^2+21 \Big] \nonumber \\
&+& \frac{  {\rm Log}[\frac{M_2^2}{\mu^2}] r_2^2 }{64 \pi ^2 (r_2^2 - 1)^5 \mu ^2} \Big[  \left(6 r_2^4+17 r_2^2+9\right) c_{4 \beta }   \nonumber \\
 && \hspace{2em}  +2 \left(r_2^4-31 r_2^2-34\right) r_2 s_{2 \beta }+r_2^6-8 r_2^4-80 r_2^2-9 \Big]
\bigg\}, \\
c_T &=& g'^4 \bigg\{  \frac{\left(17 r_1^4-7 r_1^2-4\right) c_{2 \beta }^2 }{768 \pi ^2 (r_1^2 - 1)^3 \mu ^2}-\frac{r_1^2 \left(r_1^4+2 r_1^2-2\right) c_{2 \beta }^2 {\rm Log}[\frac{M_1^2}{\mu^2}]  }{128 \pi ^2 (r_1^2 - 1)^4 \mu ^2} \bigg\}
 \nonumber \\
&+& g^4 \bigg\{ \frac{\left(29 r_2^4-7 r_2^2-16\right) c_{2 \beta }^2}{768 \pi ^2 (r_2^2 - 1)^3 \mu ^2}-\frac{r_2^2 \left(r_2^4+6 r_2^2-6\right) c_{2 \beta }^2  {\rm Log}[\frac{M_2^2}{\mu^2}]}{128 \pi ^2 (r_2^2 - 1)^4 \mu ^2}   \bigg\}
 \nonumber \\
&+& g'^2 g^2 \bigg\{   \frac{r_1^3 {\rm Log}[\frac{M_1^2}{\mu^2}]}{128 \pi ^2 (r_1^2 - 1)^4 (r_1 - r_2)^3 (r_1 + r_2)\mu ^2} \Big[ -4 c_{4 \beta }  \Big( 2 (r_2^2 - 1) r_1   \nonumber \\
 && \hspace{2em}  +r_1^5+3 r_2 r_1^4+\left(2-3 r_2^2\right) r_1^3+\left(r_2^2-8\right) r_2 r_1^2+4 r_2\Big) \nonumber \\
 && \hspace{2em} +2 \Big(-(r_2^2 - 1) r_1^4+(r_2^2 -5) r_2 r_1^3+3 r_1^6+r_2 r_1^5-\left(5 r_2^2+2\right) r_1^2  \nonumber \\
 && \hspace{2em} +\left(r_2^2+2\right) r_2 r_1+4 r_2^2\Big) s_{2 \beta }  +3 r_1^7-r_2^2 r_1^5+r_1^5 +2 r_2^3 r_1^4+r_2 r_1^4  \nonumber \\
 && \hspace{2em} -7 r_2^2 r_1^3+r_1^3-r_2^3 r_1^2-8 r_2 r_1^2+5 r_2^2 r_1-2 r_1+2 r_2^3+4 r_2 \Big] \nonumber \\
&-& \frac{r_2^3 {\rm Log}[\frac{M_2^2}{\mu^2}]}{128 \pi ^2 (r_2^2 - 1)^4 (r_1 - r_2)^3 (r_1 + r_2)\mu ^2} \Big[  - c_{4 \beta } \Big( (r_2^2 r_1^3+\left(2 r_2-3 r_2^3\right) r_1^2  \nonumber \\
 && \hspace{2em} +\left(3 r_2^4-8 r_2^2+4\right) r_1+\left(r_2^4+2 r_2^2-2\right) r_2\Big)  \nonumber \\
 && \hspace{2em} + 2 \Big(\left(r_2^3+r_2\right) r_1^3-\left(r_2^4+5 r_2^2-4\right) r_1^2+\left(r_2^4-5 r_2^2+2\right) r_2 r_1  \nonumber \\
 && \hspace{2em} +\left(3 r_2^4+r_2^2-2\right) r_2^2\Big) s_{2 \beta } +3 r_2^7-r_1^2 r_2^5+r_2^5  +2 r_1^3 r_2^4+r_1 r_2^4 \nonumber \\
 && \hspace{2em} -7 r_1^2 r_2^3+r_2^3-r_1^3 r_2^2-8 r_1 r_2^2+5 r_1^2 r_2-2 r_2+2 r_1^3+4 r_1 \Big] \nonumber \\
&+& \frac{1}{768 \pi ^2 (r_1^2 - 1)^3 (r_2^2 - 1)^3 (r_1 - r_2)^2 \mu ^2}  \Big[ \Big( \left(-2 r_2^5-5 r_2^3+r_2\right) r_1^7  \nonumber \\
 && \hspace{2em} +\left(4 r_2^6+17 r_2^4-19 r_2^2+4\right) r_1^6 -r_2 \left(2 r_2^6+12 r_2^4-57 r_2^2+25\right) r_1^5    \nonumber \\
 && \hspace{2em} +\left(17 r_2^6-126 r_2^4+120 r_2^2-29\right) r_1^4  +\left(-5 r_2^7+57 r_2^5-126 r_2^3+56 r_2\right) r_1^3   \nonumber \\
 && \hspace{2em} +\left(-19 r_2^6+120 r_2^4-102 r_2^2 +19\right) r_1^2+\left(r_2^6-25 r_2^4+56 r_2^2-26\right) r_2 r_1  \nonumber \\
 && \hspace{2em} +\left(4 r_2^4-29 r_2^2+19\right) r_2^2\Big) c_{4 \beta }  \nonumber \\
 && \hspace{2em} +2 \Big( \left(13 r_2^4-2 r_2^2+1\right) r_1^7+\left(-25 r_2^5+2 r_2^3+11 r_2\right) r_1^6  \nonumber \\
 && \hspace{2em} +\left(-25 r_2^6+48 r_2^4-93 r_2^2+34\right) r_1^5  +\left(13 r_2^6+48 r_2^4+9 r_2^2-34\right) r_2 r_1^4  \nonumber \\
 && \hspace{2em} +\left(2 r_2^6+9 r_2^4+48 r_2^2-23\right) r_1^3  +\left(-2 r_2^7-93 r_2^5+48 r_2^3+11 r_2\right) r_1^2  \nonumber \\
 && \hspace{2em} +\left(11 r_2^4-34 r_2^2+11\right) r_2^2 r_1+\left(r_2^4+34 r_2^2-23\right) r_2^3\Big)  s_{2 \beta }   \nonumber \\
 && \hspace{2em} + 3\Big(3 r_2 r_1^5 \left((r_2^2 - 1)+2 r_2^6-8 r_2^4\right)+r_2^2 \left((r_2^2 -5)+10 r_2^4\right)  \nonumber \\
 && \hspace{2em} +3 \left(2 r_2^5-r_2^3+r_2\right) r_1^7+\left(-20 r_2^6+29 r_2^4-25 r_2^2+10\right) r_1^6  \nonumber \\
 && \hspace{2em}  +\left(29 r_2^6+6 r_2^4-18 r_2^2+1\right) r_1^4 -3 r_2 \left(r_2^6-r_2^4-10 r_2^2+4\right) r_1^3  \nonumber \\
 && \hspace{2em} -\left(25 r_2^6+18 r_2^4-30 r_2^2+5\right) r_1^2+3 \left(r_2^6-r_2^4-4 r_2^2+2\right) r_2 r_1\Big) \Big]
\bigg\}.
\end{eqnarray}


\begin{thebibliography}{99}

\bibitem{Henning:2014wua}
  B.~Henning, X.~Lu and H.~Murayama,
  ``How to use the Standard Model effective field theory,''
  JHEP {\bf 1601}, 023 (2016)
  [arXiv:1412.1837 [hep-ph]].


\bibitem{Brivio:2017vri}
  I.~Brivio and M.~Trott,
  ``The Standard Model as an Effective Field Theory,''
  arXiv:1706.08945 [hep-ph].


\bibitem{Buchmuller:1985jz}
  W.~Buchmuller and D.~Wyler,
  ``Effective Lagrangian Analysis of New Interactions and Flavor Conservation,''
  Nucl.\ Phys.\ B {\bf 268}, 621 (1986).


\bibitem{Grzadkowski:2010es}
  B.~Grzadkowski, M.~Iskrzynski, M.~Misiak and J.~Rosiek,
  ``Dimension-Six Terms in the Standard Model Lagrangian,''
  JHEP {\bf 1010}, 085 (2010)
  [arXiv:1008.4884 [hep-ph]].


\bibitem{Gaillard:1985uh}
  M.~K.~Gaillard,
  ``The Effective One Loop Lagrangian With Derivative Couplings,''
  Nucl.\ Phys.\ B {\bf 268}, 669 (1986).

\bibitem{Chan:1986jq}
  L.~H.~Chan,
  ``Derivative Expansion for the One Loop Effective Actions With Internal Symmetry,''
  Phys.\ Rev.\ Lett.\  {\bf 57}, 1199 (1986).

\bibitem{Cheyette:1987qz}
  O.~Cheyette,
  ``Effective Action for the Standard Model With Large Higgs Mass,''
  Nucl.\ Phys.\ B {\bf 297}, 183 (1988).


\bibitem{Drozd:2015rsp}
  A.~Drozd, J.~Ellis, J.~Quevillon and T.~You,
  ``The Universal One-Loop Effective Action,''
  JHEP {\bf 1603}, 180 (2016)
  [arXiv:1512.03003 [hep-ph]].


\bibitem{Ellis:2016enq}
  S.~A.~R.~Ellis, J.~Quevillon, T.~You and Z.~Zhang,
  ``Mixed heavy?light matching in the Universal One-Loop Effective Action,''
  Phys.\ Lett.\ B {\bf 762}, 166 (2016)
  [arXiv:1604.02445 [hep-ph]].


\bibitem{Henning:2016lyp}
  B.~Henning, X.~Lu and H.~Murayama,
  ``One-loop Matching and Running with Covariant Derivative Expansion,''
  JHEP {\bf 1801}, 123 (2018)
  [arXiv:1604.01019 [hep-ph]].


\bibitem{Zhang:2016pja}
  Z.~Zhang,
  ``Covariant diagrams for one-loop matching,''
  JHEP {\bf 1705}, 152 (2017)
  [arXiv:1610.00710 [hep-ph]].


\bibitem{delAguila:2016zcb}
  F.~del Aguila, Z.~Kunszt and J.~Santiago,
  ``One-loop effective lagrangians after matching,''
  Eur.\ Phys.\ J.\ C {\bf 76}, no. 5, 244 (2016)
  [arXiv:1602.00126 [hep-ph]].


\bibitem{Fuentes-Martin:2016uol}
  J.~Fuentes-Martin, J.~Portoles and P.~Ruiz-Femenia,
  ``Integrating out heavy particles with functional methods: a simplified framework,''
  JHEP {\bf 1609}, 156 (2016)
  [arXiv:1607.02142 [hep-ph]].

\bibitem{Ellis:2017jns}
  S.~A.~R.~Ellis, J.~Quevillon, T.~You and Z.~Zhang,
  ``Extending the Universal One-Loop Effective Action: Heavy-Light Coefficients,''
  JHEP {\bf 1708}, 054 (2017)
  [arXiv:1706.07765 [hep-ph]].


\bibitem{Huo:2015nka}
  R.~Huo,
  ``Effective Field Theory of Integrating out Sfermions in the MSSM: Complete One-Loop Analysis,''
  Phys.\ Rev.\ D {\bf 97}, no. 7, 075013 (2018)
  [arXiv:1509.05942 [hep-ph]].


\bibitem{Huo:2015exa}
  R.~Huo,
  ``Standard Model Effective Field Theory: Integrating out Vector-Like Fermions,''
  JHEP {\bf 1509}, 037 (2015)
  [arXiv:1506.00840 [hep-ph]].


\bibitem{Chiang:2015ura}
  C.~W.~Chiang and R.~Huo,
  ``Standard Model Effective Field Theory: Integrating out a Generic Scalar,''
  JHEP {\bf 1509}, 152 (2015)
  [arXiv:1505.06334 [hep-ph]].


\bibitem{Henning:2014gca}
  B.~Henning, X.~Lu and H.~Murayama,
  ``What do precision Higgs measurements buy us?,''
  arXiv:1404.1058 [hep-ph].




\bibitem{Wells:2017vla}
  J.~D.~Wells and Z.~Zhang,
  ``Effective field theory approach to trans-TeV supersymmetry: covariant matching, Yukawa unification and Higgs couplings,''
  arXiv:1711.04774 [hep-ph].

\bibitem{Cao:2017oez}
  Q.~H.~Cao, F.~P.~Huang, K.~P.~Xie and X.~Zhang,
  ``Testing the electroweak phase transition in scalar extension models at lepton colliders,''
  Chin.\ Phys.\ C {\bf 42}, no. 2, 023103 (2018)
  [arXiv:1708.04737 [hep-ph]].



\bibitem{Huang:2016odd}
  F.~P.~Huang, Y.~Wan, D.~G.~Wang, Y.~F.~Cai and X.~Zhang,
  ``Hearing the echoes of electroweak baryogenesis with gravitational wave detectors,''
  Phys.\ Rev.\ D {\bf 94}, no. 4, 041702 (2016)
  [arXiv:1601.01640 [hep-ph]].


\bibitem{Gan:2017mcv}
  X.~Gan, A.~J.~Long and L.~T.~Wang,
  ``Electroweak sphaleron with dimension-six operators,''
  Phys.\ Rev.\ D {\bf 96}, no. 11, 115018 (2017)
  [arXiv:1708.03061 [hep-ph]].


\bibitem{Ellis:2017kfi}
  J.~Ellis, P.~Roloff, V.~Sanz and T.~You,
  ``Dimension-6 Operator Analysis of the CLIC Sensitivity to New Physics,''
  JHEP {\bf 1705}, 096 (2017)
  [arXiv:1701.04804 [hep-ph]].


\bibitem{Ferreira:2016jea}
  F.~Ferreira, B.~Fuks, V.~Sanz and D.~Sengupta,
  ``Probing ${CP}$-violating Higgs and gauge-boson couplings in the Standard Model effective field theory,''
  Eur.\ Phys.\ J.\ C {\bf 77}, no. 10, 675 (2017)
  [arXiv:1612.01808 [hep-ph]].


\bibitem{Liu:2016gzs}
  Z.~Liu,
  ``Probing the Higgs with angular observables at future $e^+e^?$ colliders,''
  Int.\ J.\ Mod.\ Phys.\ A {\bf 31}, no. 33, 1644005 (2016).


\bibitem{Craig:2015wwr}
  N.~Craig, J.~Gu, Z.~Liu and K.~Wang,
  ``Beyond Higgs Couplings: Probing the Higgs with Angular Observables at Future e$^{+}$ e$^{?}$ Colliders,''
  JHEP {\bf 1603}, 050 (2016)
  [arXiv:1512.06877 [hep-ph]].


\bibitem{Liu:2016idz}
  D.~Liu, A.~Pomarol, R.~Rattazzi and F.~Riva,
  ``Patterns of Strong Coupling for LHC Searches,''
  JHEP {\bf 1611}, 141 (2016)
  [arXiv:1603.03064 [hep-ph]].


%
%

\bibitem{Martin:1997ns}
  S.~P.~Martin,
  ``A Supersymmetry primer,''
  Adv.\ Ser.\ Direct.\ High Energy Phys.\  {\bf 21}, 1 (2010)
  [Adv.\ Ser.\ Direct.\ High Energy Phys.\  {\bf 18}, 1 (1998)]
  [hep-ph/9709356].

\bibitem{Ellis:1975ap}
  J.~R.~Ellis, M.~K.~Gaillard and D.~V.~Nanopoulos,
  Nucl.\ Phys.\ B {\bf 106}, 292 (1976).
  doi:10.1016/0550-3213(76)90382-5

\bibitem{Shifman:1979eb}
  M.~A.~Shifman, A.~I.~Vainshtein, M.~B.~Voloshin and V.~I.~Zakharov,
  Sov.\ J.\ Nucl.\ Phys.\  {\bf 30}, 711 (1979)
  [Yad.\ Fiz.\  {\bf 30}, 1368 (1979)].

\bibitem{Carena:2012xa}
  M.~Carena, I.~Low and C.~E.~M.~Wagner,
  JHEP {\bf 1208}, 060 (2012)
  doi:10.1007/JHEP08(2012)060
  [arXiv:1206.1082 [hep-ph]].

\bibitem{Huo:2013fga}
  R.~Huo,
  ``Electroweak Baryogenesis with a Supersymmetric Sector,''
  arXiv:1305.1973 [hep-ph].


\bibitem{Mertig:1990an}
  R.~Mertig, M.~Bohm and A.~Denner,
  Comput.\ Phys.\ Commun.\  {\bf 64}, 345 (1991).


\bibitem{Shtabovenko:2016sxi}
  V.~Shtabovenko, R.~Mertig and F.~Orellana,
  Comput.\ Phys.\ Commun.\  {\bf 207}, 432 (2016)
  [arXiv:1601.01167 [hep-ph]].



\bibitem{Hahn:1998yk}
  T.~Hahn and M.~Perez-Victoria,
  Comput.\ Phys.\ Commun.\  {\bf 118}, 153 (1999)
  [hep-ph/9807565].





\bibitem{Drozd:2015kva}
  A.~Drozd, J.~Ellis, J.~Quevillon and T.~You,
  ``Comparing EFT and Exact One-Loop Analyses of Non-Degenerate Stops,''
  JHEP {\bf 1506}, 028 (2015)
  [arXiv:1504.02409 [hep-ph]].





\bibitem{Baak:2014ora}
  M.~Baak {\it et al.} [Gfitter Group],
  ``The global electroweak fit at NNLO and prospects for the LHC and ILC,''
  Eur.\ Phys.\ J.\ C {\bf 74}, 3046 (2014)
  [arXiv:1407.3792 [hep-ph]].


\bibitem{Baer:2013cma}
  H.~Baer {\it et al.},
  ``The International Linear Collider Technical Design Report - Volume 2: Physics,''
  arXiv:1306.6352 [hep-ph].


\bibitem{AguilarSaavedra:2001rg}
  J.~A.~Aguilar-Saavedra {\it et al.} [ECFA/DESY LC Physics Working Group],
  ``TESLA: The Superconducting electron positron linear collider with an integrated x-ray laser laboratory. Technical design report. Part 3. Physics at an e+ e- linear collider,''
  hep-ph/0106315.


\bibitem{Fan:2014vta}
  J.~Fan, M.~Reece and L.~T.~Wang,
  ``Possible Futures of Electroweak Precision: ILC, FCC-ee, and CEPC,''
  JHEP {\bf 1509}, 196 (2015)
  [arXiv:1411.1054 [hep-ph]].


\bibitem{Bian:2015zha}
  L.~Bian, J.~Shu and Y.~Zhang,
  ``Prospects for Triple Gauge Coupling Measurements at Future Lepton Colliders and the 14 TeV LHC,''
  JHEP {\bf 1509}, 206 (2015)
  [arXiv:1507.02238 [hep-ph]].

\bibitem{TeraZ}
  S. Mishima, talk given at the Sixth TLEP Workshop at CERN, October 16-18, 2013.

\bibitem{Gomez-Ceballos:2013zzn}
  M.~Bicer {\it et al.} [TLEP Design Study Working Group],
  ``First Look at the Physics Case of TLEP,''
  JHEP {\bf 1401}, 164 (2014)
  [arXiv:1308.6176 [hep-ex]].




\bibitem{CMS:2017fij}
  CMS Collaboration [CMS Collaboration],
  ``Search for new physics in events with two low momentum opposite-sign leptons and missing transverse energy at $\sqrt{s}=13~\mathrm{TeV}$,''
  CMS-PAS-SUS-16-048.

\bibitem{Cheyette:1985ue}
  O.~Cheyette,
  ``Derivative Expansion of the Effective Action,''
  Phys.\ Rev.\ Lett.\  {\bf 55}, 2394 (1985).



\bibitem{Zuk:1985sw}
  J.~A.~Zuk,
  Phys.\ Rev.\ D {\bf 32}, 2653 (1985).



\end{thebibliography}
\end{document}